\documentclass[12pt]{article}

\usepackage{indentfirst}%
\usepackage{times,epsfig,graphics,amssymb,latexsym,amsmath,setspace,epstopdf,grffile,mathrsfs}
\usepackage{array,threeparttable,hyperref,url,bm}
\usepackage[usenames]{color}
\usepackage{graphicx,subfigure}
\usepackage[authoryear,sort]{natbib}
\usepackage{enumerate}
\usepackage{multirow}
\usepackage{algorithmic}
\usepackage{algorithm}
\usepackage{verbatim}
\allowdisplaybreaks

\usepackage{ntheorem}

\newtheorem{defi}{Definition}
\newtheorem{rmk}{Remark}
\newtheorem{prp}{Proposition}

\newtheorem{thm}{Theorem}
\newtheorem{coro}{Corollary}
\newtheorem{ass}{Assumption}
\theoremstyle{empty}

\makeatletter
\newcommand{\rmnum}[1]{\romannumeral #1}
\newcommand{\Rmnum}[1]{\expandafter\@slowromancap\romannumeral #1@}
\makeatother

\def\wt{\widetilde}

\def\wh{\widehat}

\def\boxit#1{\vbox{\hrule\hbox{\vrule\kern6pt
          \vbox{\kern6pt#1\kern6pt}\kern6pt\vrule}\hrule}}

\def\bse{\begin{eqnarray*}}
\def\ese{\end{eqnarray*}}
\def\be{\begin{eqnarray}}
\def\ee{\end{eqnarray}}
\def\bq{\begin{equation}}
\def\eq{\end{equation}}

\def\wh{\widehat}

\newtheorem{proposition}{Proposition}

\newcommand{\blem}{\begin{lemma}}
\newcommand{\elem}{\end{lemma}}
\newcommand{\bthe}{\begin{theorem}}
\newcommand{\ethe}{\end{theorem}}

\newtheorem{definition}{Definition}[section]
\newtheorem{lemma}[definition]{Lemma}
\newtheorem{theorem}[definition]{Theorem}

\def\delete#1{\iffalse #1 \fi}

\def\bse{\begin{eqnarray*}}
\def\ese{\end{eqnarray*}}
\def\bee{\begin{enumerate}}
\def\eee{\end{enumerate}}
\def\bqe{\begin{eqnarray}}
\def\eqe{\end{eqnarray}}
\def\bed{\begin{description}}
\def\eed{\end{description}}
\def\bei{\begin{itemize}}
\def\eei{\end{itemize}}
\def\bq{\begin{equation}}
\def\eq{\end{equation}}

\def\pmb#1{\setbox0=\hbox{#1}%
    \kern-.025em\copy0\kern-\wd0
    \kern.05em\copy0\kern-\wd0
    \kern-.025em\raise.0433em\box0 }
\def\pmbh#1#2{\setbox0=\hbox{#1}%
    \setbox1=\hbox{#2}%
    \kern-.025em\copy0\kern-\wd0
    \kern.05em\copy1\kern-\wd0
    \kern-.025em\raise.0433em\box0 }

\def\frac#1#2{{#1\over#2}}

\def\boxit#1{\vbox{\hrule\hbox{\vrule\kern6pt
   \vbox{\kern6pt#1\kern6pt}\kern6pt\vrule}\hrule}}

\def\listing#1{\vskip 4mm\begin{verbatim}\input#1 \vskip 4mm}
\def\thick#1{\hbox{\rlap{$#1$}\kern0.25pt\rlap{$#1$}\kern0.25pt$#1$}}

\def\wt{\widetilde}
\def\wh{\widehat}

\def\pmbh{{\pmb h}}

\renewcommand\today{\ifcase\month\or
   Jan\or Feb\or Mar\or Apr\or May\or
   Jun\or Jul\or Aug\or Sep\or Oct\or Nov\or
   Dec\fi
   \space\number\day, \number\year}

\newcommand{\va}{{\bm a}}

\newcommand{\vW}{{\bm W}}

\newcommand{\vx}{{\bm x}}

\newcommand{\vA}{{\bm A}}
\newcommand{\vB}{{\bm B}}

\newcommand{\bay}{\begin{array}}
\newcommand{\eay}{\end{array}}

\newcommand{\bqa}{\begin{eqnarray*}}
\newcommand{\eqa}{\end{eqnarray*}}
\newcommand{\bqan}{\begin{eqnarray}}
\newcommand{\eqan}{\end{eqnarray}}
\newcommand{\bqt}{\begin{quote}}
\newcommand{\eqt}{\end{quote}}
\newcommand{\bt}{\begin{tabbing}}
\newcommand{\et}{\end{tabbing}}
\newcommand{\bit}{\begin{itemize}}
\newcommand{\eit}{\end{itemize}}
\newcommand{\ben}{\begin{enumerate}}
\newcommand{\een}{\end{enumerate}}
\newcommand{\beq}{\begin{equation}}
\newcommand{\eeq}{\end{equation}}
\newcommand{\bdefi}{\begin{definition}}
\newcommand{\edefi}{\end{definition}}
\newcommand{\bpro}{\begin{proposition}}
\newcommand{\epro}{\end{proposition}}
\newcommand{\bco}{\begin{corollary}}
\newcommand{\eco}{\end{corollary}}
\newcommand{\bdes}{\begin{description}}
\newcommand{\edes}{\end{description}}

\def\boxit#1{\vbox{\hrule\hbox{\vrule\kern6pt\vbox{\kern6pt#1\kern6pt}\kern6pt\vrule}\hrule}}

\catcode`@=11
\def\@evenhead{\vbox{\hbox to\textwidth{\tiny \hfill \hfill \today } }}
\def\@oddhead{\vbox{\hbox to \textwidth{\tiny \hfill \hfill \today } }}

\def\tit.arg{Robust Tensor Regression with Nonconvexity: Algorithmic and Statistical Theory\footnote{This work was supported in part by the National Social Science Fund (22BTJ025), the Postgraduate Research \& Practice Innovation Program of Jiangsu Province (KYCX24\_3622) and the Humanities and Social Sciences Youth Foundation of Ministry of Education of China (23YJC910003).}}
\def\author.arg{Zihao Song$^a$, Jicai Liu$^b$, Heng Lian$^{c,d}$, Weihua Zhao$^{a}$\footnote{Corresponding author: Weihua Zhao, E-mail, zhaowhstat@163.com}\\
a. School of Mathematics and Statistics,  Nantong University, Jiangsu Nantong, 226019, China\\
b. School of Statistics and Mathematics\\ Shanghai Lixin University of Accounting and Finance, Shanghai, 201620, China\\
c. Department of Mathematics, City University of Hong Kong, Hong Kong, China\\
d.  City University of Hong Kong Shenzhen Research Institute, Shenzhen, 518057, China}

\begin{document}
\pagenumbering{arabic}
\setcounter{page}{1}
\baselineskip=18pt

\begin{center}
{\Large \tit.arg} \\
\vskip 3mm
\author.arg
\end{center}

\vskip 3mm

\date{}
\begin{abstract}
 Tensor regression is an important tool for tensor data analysis, but existing works have not considered the impact of outliers, making them potentially sensitive to such data points. This paper proposes a low tubal rank robust regression method for analyzing high-dimensional tensor data with heavy-tailed random noise. The proposed method is based on a nonconvex relaxation of the tensor tubal rank within a general optimization framework, which allows for nonconvexity in both the loss and penalty functions. We  develop  an implementable estimation algorithm and establish  its global convergence under some mild assumptions. Furthermore, we provide general statistical theories regarding stationary point, including the rates of convergence and bounds on the prediction error. These theoretical results cover many important models, such as linear models, generalized linear models, and Huber regression, and even encompass some nonconvex losses like correntropy and minimum distance criterion-induced losses. Supportive numerical evidence is provided through simulations and application studies.

\end{abstract}
\par

\section{Introduction}\label{Sec1}

With the rapid advancement of modern science and technology, diverse and complex data structures are prevalent across a variety of fields, including neuroimaging \citep{doi:10.1080/01621459.2013.776499}, finance \citep{doi:10.1080/01621459.2022.2063131}, computer vision \citep{10078018} and many others. These complex datasets often consist of multi-dimensional structural information, which traditional models for vector and matrix-valued data may not accurately capture and interpret the underlying relationships among the data.  Tensors serve as a natural and essential representation of these 
multi-dimensional data, fully preserving intricate structural information, such as interactions across multiple dimensions. In recent years, tensor data analysis has received increasing attention. For example, Zhou et al. \citep{doi:10.1080/01621459.2013.776499} proposed tensor regression for neuroimaging data analysis, and Lu et al. \citep{JMLR:v21:20-383} considered quantile regression which takes tensors as predictors. 

For tensor data analysis,  it is essential to capture low-rank structures. Hence, a variety of tensor decomposition approaches and their corresponding rank functions have been proposed, such as CANDECOMP/PARAFAC (CP) decomposition  and  CP rank, Tucker decomposition and Tucker rank \citep{10.1137/07070111X},  tensor-tensor product (t-product) with its derived tensor singular value decomposition (t-SVD) and tubal rank \citep{Kilmer2011}, and so on. However, the CP rank is NP-hard to estimate, and the Tucker rank is actually multilinear rank constructed by matricization which is relatively harmful to intrinsic information within tensor structure. To this end, t-product and t-SVD \citep{Kilmer2011} consider operation and decomposition in tensor form which could preserve structural information and have been popular in various fields \citep{Roy2022EJS,10078018,LuTRPCA2020}.

As the convex surrogate of tubal rank, tensor tubal nuclear norm (t-TNN, \citep{LuTRPCA2020})  attracts more and more attention in many fields. However, t-TNN still has some shortcomings, primarily in the representation of a loose relaxation of the tubal rank function, which results in a significant gap between tubal rank minimization and the solutions obtained, yielding suboptimal results. This problem will become more serious if there are outliers in the observations. To address these issues, we borrow inspiration from nonconvex regularizers used in sparse learning \citep{Fan2009ASO,SCAD01,Zhang2011AGT} and develop the exploration of nonconvex surrogates for tubal rank. These nonconvex penalties have been extensively studied, demonstrating their superiority over convex penalties, such as $l_1$ and matrix nuclear norm \citep{Cands2007EnhancingSB,NNL15TIP,Gui2015TowardsFR}. Due to the complex structure of t-SVD, it is nontrivial to develop nonconvex surrogates for tubal rank. More importantly, to our knowledge, the statistical properties of nonconvex tensor regression with tubal rank have not been previously explored, and our research aims to fill this gap to obtain more efficient estimator. 

The basic way to conquer the negative effect of the existing heavy-tailed noise and/or outlier for the robustness improvements of estimation is robust loss function. Huber loss was introduced \citep{Huber73} as a convex and robust alternative to squared loss. Based on Huber loss, Sun et al. \citep{AHR19} developed adaptive Huber regression, which adjusts the robustification parameter according to the sample size, dimensionality, and moment of noise. There are many robust loss functions being proposed, while most of them are nonconvex. However, to best of our knowledge, in the realm of tensor regression, most of works \citep{Li2022RobustLT,JMLR:v21:20-383,Lu2023StatisticalPO} focus on convex loss, such as Huber loss and/or quantile loss, for robust estimation, 
ignoring some nonconvex robust losses, albeit the nonconvex loss may outperform the convex loss. For example, the correntropy induced loss (C-loss) \citep{Feng2014RobustLR}, robust logistic regression \citep{Chi2014} proposed by the minimum distance criterion \citep{Scott2009L2E}, and so on. To this end, in this work, we do not specify the exact form of loss to encompass general cases.  

\subsection{Contributions}
This paper aims to develop a general framework for tensor regression, which incorporates nonconvex loss functions and/or nonconvex regularizers, alongside an estimation algorithm designed with convergent guarantees. 
The main contributions can be summarized as follows:
\begin{enumerate}[(i)]
\item To capture the low tubal rank structure, we develop a nonconvex surrogate regularization with various nonconvex penalty functions. The rigorous theoretical analysis illustrates the proposed nonconvex regularizer offers a more accurate approximation to tubal rank function, compared to the traditional t-TNN (please see Lemma B.4).

\item A general framework for tensor regression is proposed which incorporates  nonconvex loss functions and/or regularizers. Under the weaker condition named as local restricted strong convexity, this flexibility allows for greater adaptability and robustness in tensor regression modeling for more effective solutions in practical applications. More importantly, this framework accommodates a diverse range of loss functions and penalties, irrespective of their convexity, as long as they satisfy the specified assumptions

\item A unified estimation algorithm is developed to solve the proposed general tensor regression by the reweighting strategy. The Barzilai-Borwein initialization \citep{Barzilai1988TwoPointSS} at each iteration process is utilized to accelerate convergence, and we prove that the sequence generated by the proposed algorithm decreases monotonically. Moreover, by Kurdyka-{\L}ojasiwicz (K{\L}) property \citep{KLnon,Kurdyka1988OnGO}, the global convergence and convergence rate of estimation algorithm have been established, showing that it converges with finite steps at best, linear rate for some cases, and sublinear rate at least.

\item Under some common regularity conditions, some statistical properties of our proposed tensor regression framework are carefully studied including bounds for the Frobenius norm, nuclear norm and prediction error. We further present detailed theoretical results for linear tensor regression, generalized linear models (GLMs), adaptive Huber tensor regression, the correntropy-induced loss and the minimum distance criterion-induced logistic loss for robust tensor regression. These results encompass both nonconvex penalties and/or t-TNN regularization.

\end{enumerate}

\subsection{Related works}
{\it 1) Tensor regression}. 
Zhou et al. \citep{doi:10.1080/01621459.2013.776499} and Li et al. \citep{Li2018TuckerTR}, respectively,  adopted CP and Tucker decomposition to model the effect within  attention deficit hyperactivity disorder data previously. By two different loss functions and CP decomposition, tensor ridge regression and support tensor regression \citep{5986711} have been proposed with automatic selection of the rank. Regularized tensor regressions are firstly proposed by Raskutti et al. \citep{Raskutti2015ConvexRF}, by the decomposability of SNN, and Lian \citep{9174817} further developed support tensor machine. To accelerate computation, a smoothed support tensor machine \citep{Wang2024DensityconvolutedTS} with CP decomposition have been proposed by density convolution. Then, Lu et al. \citep{JMLR:v21:20-383,Lu2023StatisticalPO} proposed tensor quantile regression using Tucker decomposition and SNN, respectively. To obtain robust estimation, Li \citep{Li2022RobustLT} adopted Huber loss and Tucker decomposition for robust tensor regression by the generalized framework of tucker tensor estimation \citep{Hanaos22}. Furthermore, for longitudinal data, Ke et al. \citep{KE2023107609} developed tensor quantile regression via smoothing technique and CP decomposition.  Llosa-Vite et al. \citep{9749865} considered tensor-on-tensor regression with variance analysis using different tensor decomposition including CP, Tucker and tensor ring. 
However, it is well-know that these decompositions all require matricization which may destroy the spatial information within tensor and lead to suboptimal results. Hence, a regularized tensor regression \citep{Roy2022EJS} is proposed by the squared loss and tubal nuclear norm, but it is sensitive to outliers due to the nature of squared loss.

{\it 2) Nonconvex regularizer}.
As mentioned above, the nonconvex regularizers are firstly proposed to reduce model bias for sparse learning \citep{SCAD01,Zhang2011AGT}. There are many evidence proving that the nonconvex penalties outperform convex methods \citep{Cands2007EnhancingSB}. Based on nonconvex regularizers, Gui et al. \citep{Gui2015TowardsFR} considered the problem of low rank matrix estimation, and Lu et al. \citep{NNL15TIP} extended them to matrix completion. Benefited from its good performance, a nature generalization to tensor tubal rank is developed for tensor completion \citep{9340243}. However, to the best of our knowledge, using nonconvex regularizers in tensor regression has not been investigated before. Moreover, the statistical properties for nonconvex regularizers of tensor tubal rank is relatively difficult due to the complex structure, which we will consider as follows.

{\it 3) Robust estimation}. 
Due to the negative effect of heavy-tailed noise, this naturally prompts us to consider robust estimation. The absolute deviation loss is first used, but its nonsmoothness yields some problems in application. Therefore, Huber loss \citep{Huber73}, combining the advantages of squared loss and absolute deviation loss, is developed. In fact, the quantile loss \citep{Koenker2007RegressionQ} is also robust to outliers, but it aims to model the conditional quantile. In addition to these convex loss, most robust losses are nonconvex, such as Tukey’s biweight loss, and their statistical theories are widely considered \citep{Lohaos2015,JMLR:v16:loh15a}. Recently, motivated by the concept of correntropy, Feng et al. \citep{MCCRJMLR} developed robust nonparametric regression. Scott \citep{Scott2009L2E} proposed a robust framework by the minimum distance criterion and the robust logistic regression \citep{Chi2014} is developed accordingly. These appealing robust methods motivate us to develop robust tensor regression. However, due to nonconvexity, some different tools are needed for theoretical development.

\subsection{Organization and notation}
We conclude this section by presenting the organization of this paper and introducing some notations. In Section \ref{Sec2}, the background of model and some assumptions are introduced.  Section \ref{Sec3} develops estimation algorithm for the proposed nonconvex program and establish some algorithmic convergence results. In Section \ref{Sec4}, we first derive some general statistical results and then give some examples of widely-used models. Extensive numerical experiments are conducted in Section \ref{Sec5} and some conclusion are given in Section \ref{Sec6}. 

In the remainder of this paper, we denote the vectors as the lowercase boldface letters $\va$, the matrices as the capital boldface letters $\vA$, the tensors as the Euler script letters $\bm{\mathcal{A}}$. Some generic positive constants are indicated with $C,c$ which may different from line to line.
We use $\|\cdot\|_F,\|\cdot\|_{\rm op}$ and $\|\cdot\|_*$ to denote the Frobenius, operator and nuclear norm of matrix. For simplicity, we indicate the $k$-th frontal slice of $\bm{\mathcal{A}}$ with $\bm{\mathcal{A}}^{(k)}$. The inner product of $\bm{\mathcal{A}}$ and $\bm{\mathcal{B}}$ is defined as $\langle\bm{\mathcal{A}},\bm{\mathcal{B}}\rangle:={\rm vec}^{\mathsf T}(\bm{\mathcal{A}}){\rm vec}(\bm{\mathcal{B}})$ where ${\rm vec}(\cdot)$ is the vectorization operation of tensor, thus the Frobenius norm of tensor can be defined as $\|\bm{\mathcal{A}}\|_F:=\sqrt{\langle\bm{\mathcal{A}},\bm{\mathcal{A}}\rangle}$. For any random variable $x$, we denote $\|x\|_{\psi}:={\rm inf}\{t>0:\mathbb{E}[\psi(\frac{|x|}{t})]\le 1 \}$ as the $\psi$-Orlicz norm where $\psi$ indicates a nondecreasing, convex function satisfying $\psi(0)=0$. If $p\ge1$ and let $\psi_p(x):=e^{x^p}-1$, the corresponding Orlicz norm is indicated with $\|x\|_{\psi_p}:={\rm inf}\{t>0:\mathbb{E}[e^{\frac{|x|^p}{t^p}}]\le 2 \}$. We have the so-called $\nu$-sub-Gaussian random variable $x$ if $\|x\|_{\psi_2}\le\sqrt{2}\nu$.
 For any two real number sequence $a_n$ and $b_n$, we write $a_n\asymp b_n$ if $a_n/b_n\rightarrow C$. The related theory about t-SVD and all the proofs are delegated in the appendix.

\section{Model and assumptions}\label{Sec2}
Before developing a general framework of regularized tensor regression, we first list some notations of model and basic conditions on the nonconvex regularization and the risk functions.
\subsection{Model}
Suppose that $\{(y_i,\bm{\mathcal{X}}_i)\}_{i=1}^n$ denote the collected samples, drawn from the population $(y,\bm{\mathcal{X}})$, where $y$ is the univariate response and $\bm{\mathcal{X}}\in\mathbb{R}^{d_1\times d_2\times d_3}$ indicates the 3-order predictor tensor. In this paper, we consider the case of 3-order tensor predictor, and our results can be easily extended to the high-order tensor with a light modification. Further, for brevity, we assume that $d_1=d_2=d$ and the results can be adapted to the case of different $d_1$ and $d_2$ by replacing $d$ with ${\rm max}(d_1,d_2)$. Notably, in this paper, we omit the intercept with the data centralized ahead.
A differentiable empirical risk function $L_n:\mathbb{R}^{d\times d\times d_3}\times \bm{\mathcal{X}}\rightarrow \mathbb{R}$, which may be nonconvex, is considered, and the value $L_n(\bm{\mathcal{B}})$  measures the goodness-of-fit between the coefficient tensor $\bm{\mathcal{B}}$ and the collected data $\{(y_i,\bm{\mathcal{X}}_i)\}_{i=1}^n$. It is known that the empirical risk function serves as a surrogate to the population risk function $L$:
$$
L(\bm{\mathcal{B}})=\mathbb{E}[L_n(\bm{\mathcal{B}})],
$$
where the expectation is with respect to the joint distribution of $(y,\bm{\mathcal{X}})$. The true parameter tensor of interest $\bm{\mathcal{B}}^0$ is assumed to be unique and obtained by minimizing the population risk:
$$
\bm{\mathcal{B}}^0:=\mathop{\rm argmin}\limits_{\bm{\mathcal{B}}\in\mathbb{R}^{d\times d\times d_3}}L(\bm{\mathcal{B}}).
$$

To estimate the coefficient tensor $\bm{\mathcal{B}}$, we consider the framework of regularized M-estimators:
\begin{equation}\label{md1}
\wh{\bm{\mathcal{B}}}\in\mathop{\rm argmin}\limits_{\bm{\mathcal{B}}\in\mathbb{R}^{d\times d\times d_3}}L_n(\bm{\mathcal{B}})+\rho_\lambda(\bm{\mathcal{B}}),
\end{equation}
where the regularization $\rho_\lambda$ is a nonconvex surrogate to the t-TNN (see Definition A.10 in Appendix A), and the positive tuning parameter $\lambda$ works to enforce the low-rank structure on the solution.  
\begin{rmk}
Taking $d_3=1$, one can see that the considered problem reduces to that regarding to matrix. Actually, the tubal rank of a 3-order tensor is equivalent to the rank of a matrix by viewing it as a matrix of size $d\times d$ where each entry is a tube along the third dimension. Hence, similar to the linear dependence among columns and rows of matrix, a dependence, a.k.a, t-linear dependence, between different dimension, is imposed by low tubal rank \citep{Kilmer2011,doi:10.1137/110837711}. It implies that the coefficient tensor $\bm{\mathcal{B}}$ of low tubal rank captures the baseline effects to characterize the relation of response and covariate tensor.
\end{rmk}

Formally speaking, the definition of such nonconvex regularization is given as follows.
\begin{defi}\label{nonCS}
Let $\bm{\mathcal{T}}=\bm{\mathcal{U}}*\bm{\mathcal{S}}*\bm{\mathcal{V}}^{\mathsf H}$ be the t-SVD (see Definition A.6  in Appendix A)  of $\bm{\mathcal{T}}\in\mathbb{R}^{d\times d\times d_3}$. We define
$$
\rho_\lambda(\bm{\mathcal{T}}):=\frac{1}{d_3}\sum_{i=1}^d\sum_{j=1}^{d_3}\rho_\lambda(\sigma_{ij}),
$$
where  $\sigma_{ij}=\wt{\bm{\mathcal{S}}}(i,i,j)$ denotes the singular value of tensor $\bm{\mathcal{T}}$ in the Fourier domain and the univariate function $\rho_\lambda:\mathbb{R}^+\rightarrow\mathbb{R}^+$ is continuous, concave and monotonically
increasing.
\end{defi}

\begin{rmk}
The above definition contains all information about basic singular values of tensor in the Fourier domain. From Lemma B.4, as expected, the nonconvex regularizers $\rho_\lambda(\bm{\mathcal{B}})$ is a tighter envelop than t-TNN. Hence, in comparison with t-TNN, $\rho_\lambda(\bm{\mathcal{B}})$ approaches to the tensor tubal rank better and yields more sparsity among tensor singular values, which further enhances estimation accuracy and reduces the bias incurred by regularization. 
\end{rmk}

It is clear that $\rho_\lambda(\bm{\mathcal{T}})$ is separable with respect to the singular value $\sigma_{ij}$. Additionally, our goal is to develop a general framework for nonconvex tensor regression, where the convexity of loss function is not necessary. But some basic conditions do require.

\subsection{Assumption on nonconvex regularizer}
We first summarize some adopted nonconvex penalty functions in Table \ref{TNCPF}, which have been widely used in the sparse learning and satisfy the required condition in Definition \ref{nonCS}. 
\begin{table}[!htb]
\begin{center}
\caption{The formulations of considered nonconvex functions.}\label{TNCPF}
\begin{tabular}{cccccc}\hline\hline
Penalty& Formulation $(\lambda>0,\gamma>1)$ \\
\hline
Geman \citep{Geman92}& $\frac{\lambda|x|}{|x|+\gamma}$\\
\hline
SCAD $(\gamma>2)$ \citep{SCAD01}& $\begin{cases}\lambda|x|,&|x|\le\lambda\\\frac{-x^2+2\gamma\lambda |x| -\lambda^2}{2(\gamma-1)},&\lambda<|x|\le\gamma\lambda\\ \frac{\lambda^2(\gamma+1)}{2},&\gamma\lambda<|x| \end{cases}  $ \\
\hline
Laplace \citep{Laplace09}&  $\lambda(1-e^{-\frac{|x|}{\gamma}})$\\
\hline
MCP \citep{MCP10}&$ \begin{cases}\lambda x-\frac{x^2}{2\gamma},&|x|\le\gamma\lambda\\ \frac{\gamma\lambda^2}{2},&\gamma\lambda<|x|   \end{cases}$\\
\hline
ETP \citep{ETP11}&$\lambda\frac{1-e^{-\gamma |x|}}{1-e^{-\gamma}}$\\
\hline
Logarithm \citep{Lograithm12} &$\lambda\frac{{\rm log}(\gamma |x|+1)}{{\rm log}(\gamma +1)}$\\
\hline\hline
\end{tabular}
\end{center}
\end{table}

\begin{figure}[!htb]
    \centering
    \includegraphics[scale=0.6]{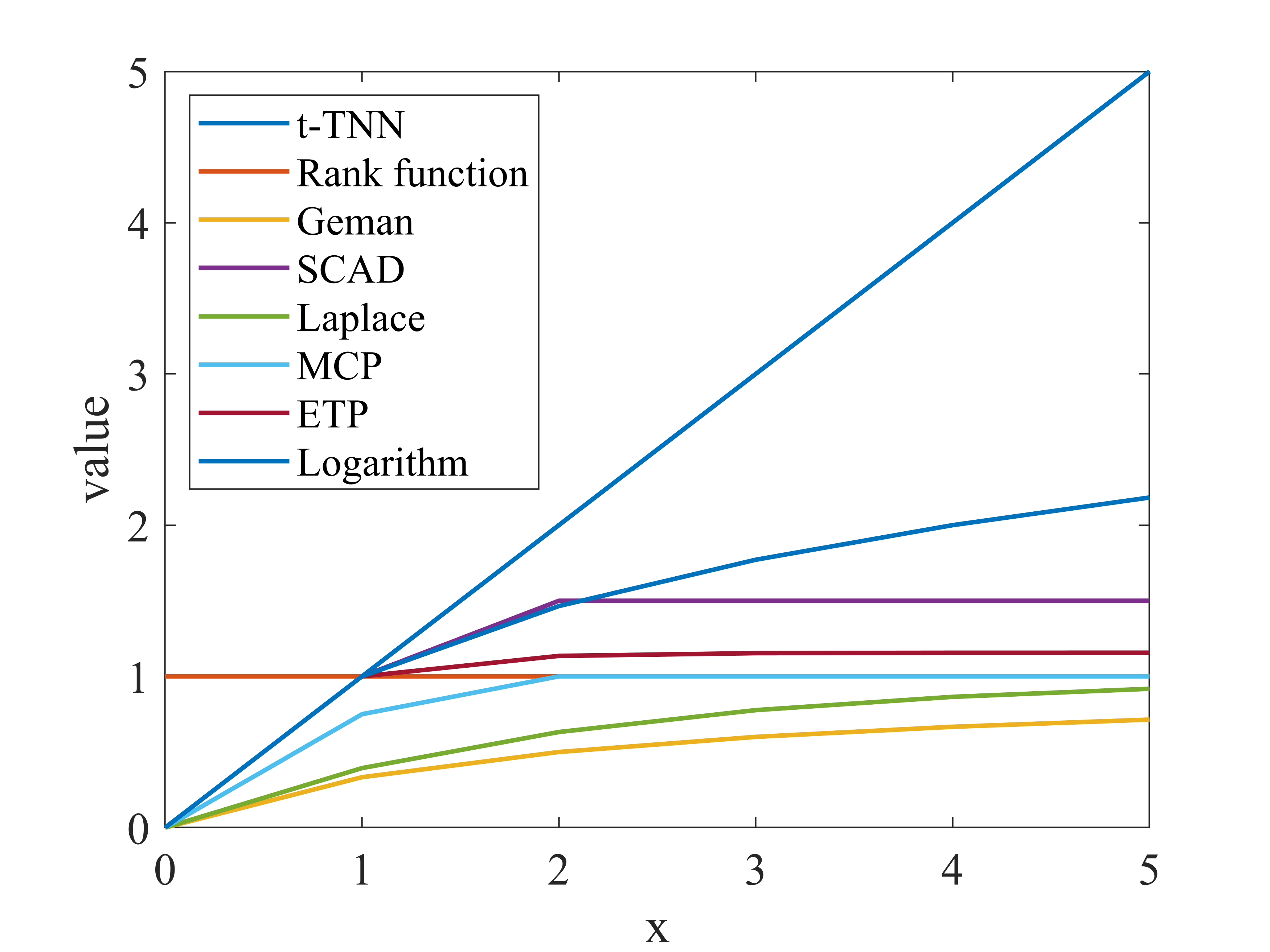}
    \caption{Plots of the t-TNN, tubal rank function and nonconvex penalties ($\lambda=1$ and $\gamma=2$).}
    \label{ncpcomFig}
\end{figure}

It is well known that these penalties play a key role to balance the gap between the ${\it l}_0$ and ${\it l}_1$ norm and achieve better performance of estimation. We extend these nonconvex penalty on the tensor singular value, which is non-trivial due to the complex structure of the tensor decomposition. In other words, the proposed nonconvex rank penalties balance the capability of rank estimation between t-TNN and tubal rank function, see Lemma B.4 for theoretical illustration and Figure \ref{ncpcomFig} for visualization. It is clear that the nonconvex penalty functions always close to t-TNN ($|x|$) for small singular values and tubal rank functions ($\mathbb{I}\{x\neq0\}$) for large singular value (see Figure \ref{ncpcomFig}), where $\mathbb{I}(\cdot)$ is the indicator function.

Compared with nonconvex counterparts, the main advantage of convex regularizers is the efficiency of computation, however, it sacrifices the convenience and performance of desired estimation and model selection \citep{Zhang2011AGT,Fan2009ASO}. For computational efficiency, in Section \ref{Sec3}, we will develop an implementable estimation algorithm with theoretical guarantee for the proposed nonconvex model. We next present some conditions that we make on the regularization terms, defined on the nonconvex function $\rho_\lambda(x)$.

\begin{ass}\label{Ass1}
~~~~~~~~~~
\begin{enumerate}[(i)]
\item The univariate function $\rho_\lambda(x)$ is symmetric around 0 and equal to $0$ at $x=0$.
\item For $x\ge 0$, $\rho_\lambda(x)$ is monotonically nondecreasing.
\item On the positive line, the function $q(x):=\frac{\rho_\lambda(x)}{x}$ is monotonically nonincreasing.
\item For all $t\neq0$, the function $\rho_\lambda$ is differentiable,  and its subdifferential at point $0$ exists with ${\rm lim}_{t\rightarrow0^+}\rho_\lambda'(x)=C_{\rho_\lambda'} $.
\item There exists $\mu>0$ such that $\rho_{\lambda,\mu}(x):=\rho_\lambda(x)+\frac{\mu}{2}x^2$ is convex.
\end{enumerate}
\end{ass}

The above assumptions are mild for the nonconvex regularizers \citep{JMLR:v16:loh15a,Lohaos2015,Zhang2011AGT} and will be used to derive statistical properties. The relatively mild assumptions (\rmnum{1})-(\rmnum{3}) can be satisfied for many penalties. Assumption (\rmnum{4}) imposes a restriction on regularizers, which excludes the ${\it l}_q$ penalty (has infinite derivative at point $0$) and the capped-${\it l}_1$ penalty (has nondifferentiable point on the positive real line). The condition (\rmnum{5}), a weak convexity describes the level of nonconvexity of penalty.

\begin{rmk}
In Appendix B, we will verify that these conditions are satisfied for our considered penalty functions in Table \ref{TNCPF}. In practice, there are many other regularization terms that are widely used satisfying the Assumption \ref{Ass1}. For instance, one can see that the standard t-TNN, $\rho_\lambda(\bm{\mathcal{B}})=\lambda\|\bm{\mathcal{B}}\|_*$, satisfies these conditions. Hence, our theoretical results can be extended to these regularizers satisfying the required conditions.
\end{rmk}


\subsection{Assumption on empirical risk}
Throughout this paper, we only require the loss function to be differentiable, without supposing it to be convex. However, for theoretical development, we impose a weaker condition, that is a type of local restricted strong convexity (LRSC).  
\begin{ass}\label{Ass2}
For a radius $\beta>0$, define the neighborhood around true parameter as $\mathscr{B}:=\{\bm{\mathcal{B}}\in\mathbb{R}^{d\times d\times d_3}:\|\bm{\mathcal{B}}-\bm{\mathcal{B}}^0\|_F\le\beta\}$. With some $\tau_1>0$ and $\tau_2\ge0$, for any $\bm{\mathcal{B}}\in\mathscr{B}$ and let $\bm{\mathcal{D}}=\bm{\mathcal{B}}-\bm{\mathcal{B}}^0$, we require
\begin{equation}\label{Ass2-LRSC}
L_n(\bm{\mathcal{B}}^0+\bm{\mathcal{D}})-L_n(\bm{\mathcal{B}}^0)-\langle L_n'(\bm{\mathcal{B}}^0) , \bm{\mathcal{D}}\rangle\ge\tau_1\|\bm{\mathcal{D}}\|_F^2-\tau_2.
\end{equation}
\end{ass}

The LRSC is defined over the neighbourhood of ground-truth parameter, different from the restricted strong convexity (RSC) used in \citep{JMLR:v16:loh15a}, where they also impose a weaker inequality outside the local region. Such condition \eqref{Ass2-LRSC} is firstly proposed by \citep{Fan2015ILAMMFS} to address sparse learning for a class of convex loss functions, and further Fan et al. \citep{FAN2019177} extends it to the generalized linear model for low-rank matrix estimation. 

\begin{rmk}
Similar to RSC, LRSC also serves for controlling a lower bound on the remainder in the first-order Taylor expansion of $L_n$.
Additionally, in the convex case, the RSC \citep{NegahbanSS2012} and LRSC \citep{FAN2019177} are typically required to hold over the intersection of a cone and a local ball of a radius centered at true parameter, where the cone is constructed by the decomposability of convex regularizers. Due to the nonconvexity, we define the LRSC only on the local ball around true parameter, and we can still derive that a similar cone condition holds by the decomposability of projection, see Lemma B.2 in appendix. 
\end{rmk}

\begin{rmk}
Note that, the condition \eqref{Ass2-LRSC} imposes no assumption on the behavior of $L_n$ outside the neighbourhood $\mathscr{B}$, which means the statistical properties of this paper focus on the local behavior of stationary points around the true parameter $\bm{\mathcal{B}}^0$, due to the nonconvexity.  Here, we remark that the radius $\beta$ could be treated as a scalar independent of sample size $n$, dimension $d,d_3$ and tubal rank $r$. The ball $\mathscr{B}$ of radius $\beta$ essentially cuts out a local region where the curvature of $L_n$ are well-behaved.
\end{rmk}

\section{Unified framework of estimation algorithm}\label{Sec3}
In this section, we propose a unified framework to optimizing the program \eqref{md1} via the iterative reweighting rule.
Albeit the objective function may be nonconvex completely, the corresponding global convergence results are still derived. Some additional assumptions are necessary due to the nonconvex nature of problem and it is of importance to develop convergence guarantee of estimation algorithm for nonconvex optimization. 

A function $f$ is called K{\L} function if the K{\L} property holds (see Definition C.1, \cite{KLnon,Kurdyka1988OnGO}) for any point in ${\rm dom}(\partial f)$.
\begin{ass}\label{Ass3}
The loss function and penalty function are K{\L} function, which implies that the objective function is K{\L} function.
\end{ass}

\begin{rmk}
Assumption \ref{Ass3} is very mild and it consists of a wide variety of functions. Specially, all semi-algebraic, log-exp and sub-analytic are K{\L} function, see \cite{Bolte13PA} and references therein.
\end{rmk}

Subsequently, we first describe the implementation of estimation algorithm by the iterative reweighting rule and then establish its convergence results.

\subsection{Estimation algorithm}
We now present an implementable estimation algorithm to obtain the desired estimator. For notation simplicity, the $k$-th iterative values are indicated with subscript $k$. Due to the concavity of the considered penalty functions in Table \ref{TNCPF}, for any $0<x<s$, we have the so-called antimonotone property \citep{NNL15TIP} as follows,
\begin{equation}\label{AoP1}
\rho_\lambda'(s)\le\frac{\rho_\lambda(x)-\rho_\lambda(s)}{x-s}\le\rho_\lambda'(x).
\end{equation}
Despite the nondifferentiable nature of $\rho_\lambda$ at some specific points, the property \eqref{AoP1} still holds and reduces to
\begin{equation}\label{AoP2}
q\le\frac{\rho_\lambda(x)-\rho_\lambda(s)}{x-s}\le p,
\end{equation}
with $q\in\partial\rho_\lambda(s) $ and $p\in\partial\rho_\lambda(x)$. Our updating strategy stems from the linearized approximation of nonconvex regularizers and the  quadratic majorization of risk function at $k$-th iteration results.

From the property \eqref{AoP2}, it implies that
$$
\rho_\lambda(\sigma_{ij})\le\rho_\lambda(\sigma_{ij,k})+w_{ij,k}(\sigma_{ij}-\sigma_{ij,k}),
$$
where $w_{ij,k}\in\partial\rho_\lambda(\sigma_{ij,k})$ are regarded as the reweighted weights based on the $\sigma_{ij,k}$ and $\sigma_{ij,k}$ are the singular values within all frontal slices of $\bm{\mathcal{B}}_k$, which leads to
\begin{equation}\label{LANP}
\rho_\lambda(\bm{\mathcal{B}})\le \frac{1}{d_3}\sum_{i=1}^d\sum_{j=1}^{d_3}\Big( \rho_\lambda(\sigma_{ij,k})+w_{ij,k}(\sigma_{ij}-\sigma_{ij,k})\Big).
\end{equation}
On the other hand, the quadratic majorization function of empirical risk $L_n$ evaluated at the $k$-th iterative estimation $\bm{\mathcal{B}}_k$ is
\begin{equation}\label{QMLF}
L_n(\bm{\mathcal{B}})= L_n(\bm{\mathcal{B}}_k)+\langle L_n'(\bm{\mathcal{B}}_k),\bm{\mathcal{B}}-\bm{\mathcal{B}}_k \rangle+\frac{\eta_k}{2}\|\bm{\mathcal{B}}-\bm{\mathcal{B}}_k\|_F^2,
\end{equation}
where $\eta_k$ is a positive constant. Furthermore, replacing the regularizer and loss function in \eqref{md1} by the right side of \eqref{LANP} and \eqref{QMLF}, respectively, so $\bm{\mathcal{B}}_{k+1}$ is updated by solving the relaxed problem:
\begin{equation}\label{SPB}
\begin{aligned}
\bm{\mathcal{B}}_{k+1}&=\mathop{\rm argmin}\limits_{\bm{\mathcal{B}}}\frac{1}{d_3}\sum_{i=1}^d\sum_{j=1}^{d_3}\Big( \rho_\lambda(\sigma_{ij,k})+w_{ij,k}(\sigma_{ij}-\sigma_{ij,k})\Big)\\
&+L_n(\bm{\mathcal{B}}_k)+\langle L_n'(\bm{\mathcal{B}}_k),\bm{\mathcal{B}}-\bm{\mathcal{B}}_k \rangle+\frac{\eta_k}{2}\|\bm{\mathcal{B}}-\bm{\mathcal{B}}_k\|_F^2
\end{aligned}
\end{equation}
Ignoring some independent constant terms, it is degraded to the weighted t-TNN (Wt-TNN, see Appendix A for its definition) proximal problems:
\begin{equation}\label{WTNNPP}
\bm{\mathcal{B}}_{k+1}=\mathop{\rm argmin}\limits_{\bm{\mathcal{B}}}\frac{1}{\eta_k}\|\bm{\mathcal{B}}\|_{\vW_k,*}+\frac{1}{2}\|\bm{\mathcal{B}}-\bm{\mathcal{G}}_k  \|_F^2
\end{equation}
where $\bm{\mathcal{G}}_k:=\bm{\mathcal{B}}_k-L_n'(\bm{\mathcal{B}}_k)/\eta_k$. Note that $\sigma_{1j,k}\ge\cdots\ge\sigma_{dj,k}\ge0$, due to the nature of $\rho_\lambda'$, we have
\begin{equation}\notag
0\le w_{1j,k}\le\cdots\le w_{dj,k},~~j=1,\cdots,d_3.
\end{equation}
Hence, the problem can by solved by the weighted tensor singular value thresholding (Wt-SVT) operator \citep{MU20204}, extended from matrix weighted SVT \citep{Chen2012ReducedRR}, that is
\begin{equation}\label{UpB}
\bm{\mathcal{B}}_{k+1}=\bm{\mathcal{U}}*\bm{\mathcal{S}}_{\vW_k,\frac{1}{\eta_k}}*\bm{\mathcal{V}}^{\mathsf T},
\end{equation}
where $$\bm{\mathcal{S}}_{\vW_k,\frac{1}{\eta_k}}={\rm ifft}((\wt{\bm{\mathcal{S}}}-\frac{1}{\eta_k}\bm{\mathcal{W}}_k)_+,[~],3),$$
$\bm{\mathcal{W}}_k\in\mathbb{R}^{d\times d\times d_3}$ is a f-diagonal tensor whose diagonal entries of the $j$-th frontal slice are identical to elements of the $j$-th column of $\vW_k$, and ${\rm ifft}(\cdot)$ denotes inverse discrete Fourier transformation in MATLAB command.

Notably, the parameter $\eta_k$ is crucial to the accuracy of the quadratic majorization \eqref{QMLF}. In addition, $\frac{1}{\eta_k}$ serves as the step size in the proximal problem \eqref{WTNNPP}. Hence, we should select $\eta_k$ carefully since it affects not only  the accuracy of \eqref{QMLF} but also the convergence behavior of solving \eqref{WTNNPP}. The popular backtracking strategy \citep{beck09} is adopted to set $\eta$, i.e., $\eta_k=\kappa\eta_k,~\kappa>1$, such that the monotonous search criterion holds
\begin{equation}\label{MSC}
\begin{aligned}
&L_n(\bm{\mathcal{B}}_{k+1})+\rho_\lambda(\bm{\mathcal{B}}_{k+1})+\frac{\alpha}{2}\eta_k\|\bm{\mathcal{B}}_{k+1}-\bm{\mathcal{B}}_{k}\|_F^2\le L_n(\bm{\mathcal{B}}_{k})+\rho_\lambda(\bm{\mathcal{B}}_{k}),
\end{aligned}
\end{equation}
where $0<\alpha<1$. The Barzilai–Borwein rule \citep{Barzilai1988TwoPointSS} is used to initialize $\eta_k$:
\begin{equation}\label{Upeta}
\eta_k:=\mathop{\rm argmin}\limits_{\eta}\|\Delta_{1k}-\frac{1}{\eta}\Delta_{2k}  \|_F^2=\frac{\langle \Delta_{1k},\Delta_{2k}   \rangle}{\langle \Delta_{1k},\Delta_{1k}   \rangle},
\end{equation}
where $\Delta_{1k}:=\bm{\mathcal{B}}_k-\bm{\mathcal{B}}_{k-1}$ and $\Delta_{2k}:=L_n'(\bm{\mathcal{B}}_k)-L_n'(\bm{\mathcal{B}}_{k-1})$. The detailed estimation procedure is delegated to Algorithm \ref{Alg1}.
\begin{algorithm}[htbp]
\caption{Estimation algorithm via iterative reweighting rule.}\label{Alg1}
\begin{algorithmic}
\STATE
\STATE \textbf{Input:} Initial value $\bm{\mathcal{B}}_0$, data $\{(y_i,\bm{\mathcal{X}}_i)\}^n_{i=1}$, $\lambda$, $\eta_0$, $\kappa>1$, $0<\alpha<1$, $\epsilon_{tol}$ and $k=0$.
\STATE \textbf{Output:} $\widehat{\bm{\mathcal{B}}}$.
\STATE {\bf while} not converge {\bf do}
\STATE 1. Update $\bm{\mathcal{B}}_{k+1}$ by
\STATE \quad {\bf repeat}
\STATE \quad\quad Compute $\bm{\mathcal{B}}_{k+1}$ by \eqref{UpB};
\STATE \quad\quad Update $\mu_{k}$ by $\eta_{k}=\kappa\eta_k$;
\STATE \quad {\bf until} the criterion \eqref{MSC} holds.
\STATE 2. Compute $\vW_{k+1}$ by $w_{ij,k}\in\partial\rho_\lambda(\sigma_{ij,k})$.
\STATE 3. {\bf if} $\|\bm{\mathcal{B}}_{k+1}-\bm{\mathcal{B}}_k\|_F\le \epsilon_{tol}$
\STATE \quad\quad {\bf break}.
\STATE 4. Let $k\leftarrow k+1$.
\STATE 5. Barzilai-Borwein Initialization: computing $\eta_{k}$ by \eqref{Upeta}.
\STATE {\bf end while}
\end{algorithmic}
\end{algorithm}

\begin{rmk}
For each iteration, we require computing FFT and SVD, hence, the per-iteration complexity is $\bm{\mathcal{O}}(d^2d_3{\rm log}(d_3)+d^3d_3)$, where the FFT costs $\bm{\mathcal{O}}(d^2d_3{\rm log}(d_3))$. For the case of matrix predictors, i.e., $d_3=1$, the per-iteration complexity reduces to $\bm{\mathcal{O}}(d^3)$.
\end{rmk}

\subsection{Convergence results}
Finally, we turn to establish the convergence properties of our proposed estimation algorithm. Denote $F(\bm{\mathcal{B}})=L_n(\bm{\mathcal{B}})+\rho_\lambda(\bm{\mathcal{B}})$ as the objective function.  Let $\{\bm{\mathcal{B}}_k,k=1,\cdots\}$ be the sequence generated by Algorithm \ref{Alg1}, and  we have the following convergence results.
\begin{thm}\label{thm-Alg1-CA1}
Suppose that the nonconvex penalty function $\rho_\lambda$ satisfies Assumption \ref{Ass1}. Assume that the parameter $\mu_k\ge C_{l}/(1-\alpha)$ where $0<\alpha<1$ for any $k>0$ and $L_n'$ is Lipschitz continuous with constant $C_{l}$, the following properties hold for the sequence $\{\bm{\mathcal{B}}_k\}$.
\begin{enumerate}[(1)]
\item $F(\bm{\mathcal{B}}_k)$ is monotonically decreasing. Furthermore, the search criterion \eqref{MSC} always holds;
\item The sequence $\{\bm{\mathcal{B}}_k\}$ satisfies  ${\rm lim}_{k\to+\infty}(\bm{\mathcal{B}}_{k+1}-\bm{\mathcal{B}}_k)=0$.
\end{enumerate}
\end{thm}

Theorem \ref{thm-Alg1-CA1} shows the generated sequence makes the value of objective function decrease monotonically and satisfies asymptotic regularity. Further, if the objective function is coercive, the local convergence result holds illustrated as Theorem \ref{thm-Alg1-CA2}.

\begin{thm}[Local convergence]\label{thm-Alg1-CA2}
Under the same assumption as Theorem \ref{thm-Alg1-CA1}. Suppose that the objective function $F$ is coercive, thus, any accumulation points $\bm{\mathcal{B}}^\star$ of $\{\bm{\mathcal{B}}_k\}$ are the stationary point of $F(\bm{\mathcal{B}})$.
\end{thm}

\begin{rmk}
The coerciveness of function $f(x)$ means that it is bounded from below and $f(x)\rightarrow\infty$ with $x\rightarrow\infty$. In Table \ref{TNCPF}, it is clear that the Logarithm penalty function is coercive. And the widely used squared loss and Huber loss \citep{Huber73} are all coercive.  In addition, it is clear that the objective function $F$ is coercive as long as one of the penalty and loss function is coercive.
\end{rmk}

\begin{thm}[Global convergence]\label{thm-Alg1-CA3}
Apart from the same conditions as Theorem \ref{thm-Alg1-CA2}, if Assumption \ref{Ass3} also holds, then the whole sequence $\{\bm{\mathcal{B}}_k\}$ converges to a stationary point of $F(\bm{\mathcal{B}})$ with a finite length, i.e., $\sum_{k=0}^{+\infty}\|\bm{\mathcal{B}}_{k+1}-\bm{\mathcal{B}}_k\|_F^2<+\infty$.
\end{thm}

Theorem \ref{thm-Alg1-CA3} shows that the sequence generated by the proposed estimation algorithm is globally convergent, although the problem is nonconvex.
For our used penalty functions in Table \ref{TNCPF}, the following Proposition \ref{PFKL} demonstrates that they are all semi-algebraic function as well as K{\L} function.
\begin{prp}\label{PFKL}
The penalty functions in Table\ref{TNCPF} are all semi-algebraic as well as K{\L} function.
\end{prp}

The following theorem shows that the convergence rate of the sequence $\{\bm{\mathcal{B}}_k\}$ is at least sublinear under the semi-algebraic assumption on the objective function.
\begin{thm}\label{thm-Alg1-CA4}
With the same conditions as Theorem \ref{thm-Alg1-CA2}, we assume that the objective function $F(\bm{\mathcal{B}})$ is semi-algebraic. Thus it satisfies the K{\L} property with taking desingularizing function $\varphi(x)=\xi x^\zeta$ in Definition C.1 where $\xi>0$ and $0<\zeta\le1$. And we have
\begin{enumerate}[(1)]
\item ({\bf Sublinear}) If $0<\zeta< 0.5$, there exists $M_1>0$ such that $\|\bm{\mathcal{B}}_{k}-\bm{\mathcal{B}}^\star\|_F\le M_1(\frac{1}{k})^{\frac{\zeta}{1-2\zeta}}$;
\item ({\bf Linear}) If $0.5\le\zeta< 1$, there exists $M_2>0$ and $0\le\varpi<1$ such that $\|\bm{\mathcal{B}}_{k}-\bm{\mathcal{B}}^\star\|_F\le M_2\varpi^k$;
\item ({\bf Finite steps}) If $\zeta=1$, thus the sequence $\{\bm{\mathcal{B}}_k\}$ converges to a stationary point $\bm{\mathcal{B}}^\star$ of $F(\bm{\mathcal{B}})$ with finite iterations.
\end{enumerate}
\end{thm}

\begin{rmk}
The determination of desingularizing function $\varphi(x)$ is still a challenging problem. Albeit there are some pleasing progress, most of them are not applicable in this paper and we refer interested readers to \cite{10.1007/s10208-017-9366-8,10.1007/s10957-023-02219-y,10.1137/20M1314057}. Specially, for MCP/SCAD penalized least square estimator, Li and Pong \citep{10.1007/s10208-017-9366-8} proved that $\zeta=0.5$, implying that the proposed algorithm enjoys linear convergence rate in this case. In general, our estimation algorithm achieve sublinear convergence rate at least. Moreover, we give some visualization on empirical convergence process, please see Figure \ref{Simu-F1} and \ref{Simu-F3} in Section \ref{Sec5}.
\end{rmk}

\section{General statistical properties and specific examples}\label{Sec4}


We next establish some statistical properties of $\wh{\bm{\mathcal{B}}}$. Some general theorems are firstly established and then we investigate statistical consistency of some examples under different loss functions.

\subsection{General results}
Before the statement of our main results, a specify condition should be imposed on the tuning parameter $\lambda$ and we will verify that it holds with high probability in following examples. Furthermore, we assume that the ground-truth parameter $\bm{\mathcal{B}}^0$ is low-rank with the tubal rank $r$.
\begin{ass}\label{Ass4}
The tuning parameter should be chosen properly such that $\lambda\ge4\|L_n'(\bm{\mathcal{B}})\|/C_{\rho_\lambda'}$.
\end{ass}
\begin{ass}\label{Ass5}
Let $\bm{\mathcal{B}}^0\in\mathop{\rm argmin}\mathbb{E}[L_n(\bm{\mathcal{B}})]$ be the true parameter values and ${\rm rank}_t(\bm{\mathcal{B}}^0)=r$.
\end{ass}

We begin with a theorem to illustrate the statistical rate of $\wh{\bm{\mathcal{B}}}$ under some regularity conditions.
\begin{thm}\label{EB1}
Under Assumption \ref{Ass1},\ref{Ass2} and \ref{Ass4},\ref{Ass5}, suppose that $3\mu<4\tau_1$, $n\ge Crdd_3/\beta^2$ and  $\tau_2= C\lambda{C_{\rho_\lambda'}}$ for some constants $C\ge0$. Let $\wh{\bm{\mathcal{B}}}$ be a stationary point of problem \eqref{md1} such that $\wh{\bm{\mathcal{B}}}\in\mathscr{B}$, then $\wh{\bm{\mathcal{B}}}$ exists and it satisfies
\begin{equation}\label{EB1-eq}
\|\wh{\bm{\mathcal{B}}}-\bm{\mathcal{B}}^0\|_F\le\frac{6\lambda C_{\rho_\lambda'} \sqrt{r}}{4\tau_1-3\mu}~~{\rm and}~~\|\wh{\bm{\mathcal{B}}}-\bm{\mathcal{B}}^0\|_*\le\frac{24\lambda C_{\rho_\lambda'} r}{4\tau_1-3\mu}.
\end{equation}
\end{thm}

\begin{rmk}
From the results of Theorem \ref{EB1}, one can see that the Frobenius norm error  grows proportionally with the tensor tubal rank $r$ and the squared values of tuning parameter  $\lambda$ . As we will see,  for many specify statistical models, we should take $\lambda\asymp\sqrt{{dd_3}/{n}}$ to satisfy the assumed conditions. In this way, the Frobenius norm error scales as $\sqrt{{rdd_3}/{n}}$ as expected. 
\end{rmk}
\begin{rmk}
The term $C_{\rho_\lambda'}$ characters the effect of the nonconvex penalty on the statistical rate. It it clear that the smaller $C_{\rho_\lambda'}$ leads to the faster statistical rate. In fact, besides the SCAD and MCP regularization, for other penalty functions, the value of $C_{\rho_\lambda'}$ is a function of the parameter $\gamma$ which controls the capability of incurring sparsity. Hence, as wanted, the bound \eqref{EB1-eq} shows the difference of various regularizers on the statistical consistency.
\end{rmk}
\begin{rmk}
It is well-known that $\tau_1$ serves as the curvature constant of the empirical risk function and $\mu$ is a measure of the nonconvexity of penalty functions. Then, $4\tau_1-3\mu$ can be regarded as a type of global measure with respect to the behavior of overall objective function \eqref{md1} and \eqref{EB1-eq} implies that the higher values (better curvature behavior) yields the faster statistical rate.
\end{rmk}

The next result gives a bound on the prediction error (PE), which is interpretable in various cases, defined as follows,
$$
{\rm PE}(\wh{\bm{\mathcal{B}}};\bm{\mathcal{B}}^0):=\langle L_n'(\wh{\bm{\mathcal{B}}})-L_n'(\bm{\mathcal{B}}^0),\wh{\bm{\mathcal{B}}}-\bm{\mathcal{B}}^0 \rangle.
$$

For instance, with the usage of widely used squared loss function, i.e.,  $L_n(\bm{\mathcal{B}})=\frac{1}{2n}\sum_{i=1}^n(y_i-\langle \bm{\mathcal{X}}_i,\bm{\mathcal{B}}\rangle)^2$, we have
$$
{\rm PE}(\wh{\bm{\mathcal{B}}};\bm{\mathcal{B}}^0)=\frac{1}{n}\sum_{i=1}^n(\langle \bm{\mathcal{X}}_i,\wh{\bm{\mathcal{B}}}-\bm{\mathcal{B}}^0 \rangle)^2,
$$
which is the common measure of the prediction error for a linear regression.
In general, for GLMs using the nonnegative log likelihood as loss function with cumulant function $\psi$, its prediction error is equivalent to the symmetrized Bregman divergence.

\begin{thm}\label{PB1}
With the same condition as Theorem \ref{EB1}, the prediction error is bounded as
\begin{equation}\notag
\begin{aligned}
{\rm PE}(\wh{\bm{\mathcal{B}}};\bm{\mathcal{B}}^0)\le\lambda^2 C_{\rho_\lambda'}^2r\Big( \frac{9}{4\tau_1-3\mu}+\frac{27\mu}{(4\tau_1-3\mu)^2}\Big).
\end{aligned}
\end{equation}
\end{thm}
The result provided above shows that the prediction error behaves similarly to the bound of squared Frobenius norm in Theorem \ref{EB1}. 
In the remainder of this section, we present some consequences, and the proof of these corollaries mainly rely on verifying the Assumption \ref{Ass2} and \ref{Ass4}, then combining Theorem \ref{EB1} and \ref{PB1} to derive results. We follow the roadmap of developing tensor linear model, GLMs and robust regression models.

\subsection{Linear tensor regression}\label{Sec-LTR}
To begin with, we consider the following ordinary linear tensor regression model,
$$
y_i=\langle \bm{\mathcal{X}}_i, \bm{\mathcal{B}}^0\rangle+\epsilon_i,~~i=1,\cdots,n,
$$
where the $\bm{\mathcal{B}}^0$ is the unknown true parameter tensor and the error term $\epsilon_i$ follows the Gaussian distribution $N(0,\sigma^2)$. Using the squared loss function, the population and empirical risk functions are then given by,
$$L(\bm{\mathcal{B}})=\mathbb{E}\Big[\frac{1}{2}(y-\langle \bm{\mathcal{X}}, \bm{\mathcal{B}}\rangle)^2\Big],~~L_n(\bm{\mathcal{B}})=\frac{1}{2n}\sum_{i=1}^n(y_i-\langle \bm{\mathcal{X}}_i, \bm{\mathcal{B}}\rangle)^2. $$
Thus, one can see that $\bm{\mathcal{B}}^0=\mathop{\rm argmin}_{\bm{\mathcal{B}}}L(\bm{\mathcal{B}})$. The desired estimator is defined by
\begin{equation}\label{LTR}
\wh{\bm{\mathcal{B}}}\in\mathop{\rm argmin}\frac{1}{2n}\sum_{i=1}^n(y_i-\langle \bm{\mathcal{X}}_i, \bm{\mathcal{B}}\rangle)^2+\rho_\lambda(\bm{\mathcal{B}}).
\end{equation}
For brevity, the Hessian of $L_n$ is denoted as $\mathbf{H}_n:=L_n''(\bm{\mathcal{B}})=\frac{1}{n}\sum_{i=1}^n\vx_i\vx_i^{\mathsf T}$ with $\vx_i={\rm vec}(\bm{\mathcal{X}}_i)$ and let $\mathbf{H}:=\mathbb{E}[\mathbf{H}_n]$.
 Then we have the following corollary.
\begin{coro}\label{LTRB}
Suppose that $\vx$ is sub-Gaussian, $3\mu<2\Lambda_{\rm min}(\mathbf{H})$, $\lambda\asymp \sqrt{dd_3/n}$ and $n\ge Crdd_3/\beta^2$. For any stationary points $\wh{\bm{\mathcal{B}}}$ of the nonconvex program \eqref{LTR}, with high probability, we have
$$
\|\wh{\bm{\mathcal{B}}}-\bm{\mathcal{B}}^0\|_F\le C\frac{C_{\rho_\lambda'}\lambda \sqrt{r}}{2\Lambda_{\rm min}(\mathbf{H})-3\mu},~~\|\wh{\bm{\mathcal{B}}}-\bm{\mathcal{B}}^0\|_*\le C\frac{C_{\rho_\lambda'}\lambda r}{2\Lambda_{\rm min}(\mathbf{H})-3\mu},
$$
and
\begin{equation}\notag
\begin{aligned}
&\frac{1}{n}\sum_{i=1}^n(\langle \bm{\mathcal{X}}_i,\wh{\bm{\mathcal{B}}}-\bm{\mathcal{B}}^0 \rangle)^2 \le C_{\rho_\lambda'}^2\lambda^2r\Big( \frac{c}{2\Lambda_{\rm min}(\mathbf{H})-3\mu}+\frac{c\mu}{(2\Lambda_{\rm min}(\mathbf{H})-3\mu)^2}\Big),
\end{aligned}
\end{equation}
where ${\rm rank}_t(\bm{\mathcal{B}}^0)=r$ and $\Lambda_{\rm min}(\mathbf{H})$ is the smallest eigenvalue of $\mathbf{H}$.
\end{coro}

\begin{rmk}
Although the LRSC \eqref{Ass2-LRSC} holds for any $\bm{\mathcal{D}}=\bm{\mathcal{B}}-\bm{\mathcal{B}}^0$ since the squared loss function is convex, the stated results of Corollary \ref{LTRB} shows the statistical consistency for a global minimum within the local ball with radius $\beta$ due to the nonconvexity of regularizers. However, by the convexity of squared loss, the radius $\beta$ can be further allowed to diverge, which implies that it is a strong result holding for any stationary point. In the end, our proposed estimation algorithm generates a sequence that converges to a stationary point approaching to the true parameter with the sublinear convergence rate at least. 
\end{rmk}


Moreover, if taking $\rho_\lambda(\bm{\mathcal{B}})=\lambda\|\bm{\mathcal{B}}\|_*$, it reduces to t-TNN and the Corollary \ref{LTRB} degrades to the following results.
\begin{prp}
Suppose that $\vx$ is sub-Gaussian, $0<\Lambda_{\rm min}(\mathbf{H})$ and $\lambda\asymp \sqrt{dd_3/n}$. With high probability, we have
$$
\|\wh{\bm{\mathcal{B}}}-\bm{\mathcal{B}}^0\|_F\le C\frac{\lambda \sqrt{r}}{\Lambda_{\rm min}(\mathbf{H})},~~\|\wh{\bm{\mathcal{B}}}-\bm{\mathcal{B}}^0\|_*\le C\frac{\lambda r}{\Lambda_{\rm min}(\mathbf{H})},
$$
and
$$
\frac{1}{n}\sum_{i=1}^n(\langle \bm{\mathcal{X}}_i,\wh{\bm{\mathcal{B}}}-\bm{\mathcal{B}}^0 \rangle)^2\le C \frac{\lambda^2 r}{\Lambda_{\rm min}(\mathbf{H})}.
$$
\end{prp}
From now on, we always omit the proof of results with respect to the t-TNN penalty, without additional statement, since it is trivial.
If we take $d_3=1$, it is clear that the results are consistent with that of matrix predictors. For instance, from \cite{FAN2019177}, we know that
\begin{equation}\notag
\|\widehat{\vB}-\vB^0\|_F\le C\frac{\lambda\sqrt{r}}{\Lambda_{\rm min}(\mathbf{H})},
\end{equation}
where $\wh{\vB}$ is the estimation of true matrix parameter $\vB^0\in\mathbb{R}^{d\times d}$ and $\lambda$ takes $\sqrt{{d}/{n}}$.

\subsection{Generalized tensor regression}\label{Sec-GLM}
Furthermore, moving beyond linear regression, we consider that the observations follows the GLMs. Suppose that the responses $y_i$ are drawn from the exponential family, that is
$$
P(y_i|\bm{\mathcal{X}}_i,\bm{\mathcal{B}})={\rm exp}\Big(\frac{y_i\langle \bm{\mathcal{X}}_i,\bm{\mathcal{B}}^0\rangle-\psi(\langle \bm{\mathcal{X}}_i,\bm{\mathcal{B}}^0\rangle)}{w}\Big),
$$
where $\psi(\cdot)$ is specified
function and $w$ is a positive constant. Disregarding some independent terms, the population and empirical risk function, corresponding to the negative log likelihood, are as follows
$$
L(\bm{\mathcal{B}})=\mathbb{E}[-y\langle \bm{\mathcal{X}},\bm{\mathcal{B}} \rangle+\psi(\langle \bm{\mathcal{X}},\bm{\mathcal{B}}\rangle)],~~L_n(\bm{\mathcal{B}})=\frac{1}{n}\sum_{i=1}^n\big( -y_i\langle \bm{\mathcal{X}}_i,\bm{\mathcal{B}} \rangle+\psi(\langle \bm{\mathcal{X}}_i,\bm{\mathcal{B}}\rangle)\big).
$$
Let
$\mathbf{H}_n(\bm{\mathcal{B}})=L_n''(\bm{\mathcal{B}} )=\frac{1}{n}\sum_{i=1}^n\psi''(\langle \bm{\mathcal{X}}_i,\bm{\mathcal{B}}\rangle)\vx_i\vx_i^{\mathsf T}$
and $\mathbf{H}(\bm{\mathcal{B}})=\mathbb{E}[\mathbf{H}_n(\bm{\mathcal{B}})]$. Then, we  assume that $\bm{\mathcal{B}}^0$ has low tubal rank $r$, and recruit the following M-estimator
\begin{equation}\label{GLMTR}
\wh{\bm{\mathcal{B}}}\in\mathop{\rm argmin} \frac{1}{n}\sum_{i=1}^n\big( \psi(\langle \bm{\mathcal{X}}_i,\bm{\mathcal{B}}\rangle)-y_i\langle \bm{\mathcal{X}}_i,\bm{\mathcal{B}} \rangle\big)+\rho_\lambda(\bm{\mathcal{B}}).
\end{equation}

\begin{coro}\label{GLMB}
Suppose that $\vx$ is sub-Gaussian, $3\mu<4\Lambda_{\rm min}(\mathbf{H}(\bm{\mathcal{B}}^0))$, $|\psi''(s)|\le c$ for any $s\in\mathbb{R}$ with a positive constant $c$ and $|\psi'''(s)|\le|s|^{-1}$ for $|s|> 1$. Given $\lambda\asymp\sqrt{dd_3/n}$ and $n\ge Crdd_3/\beta^2$, with high probability, for any stationary point $\wh{\bm{\mathcal{B}}}$ of the nonconvex problem \eqref{GLMTR}, the estimation error bounds are
$$
\|\wh{\bm{\mathcal{B}}}-\bm{\mathcal{B}}^0\|_F\le C\frac{C_{\rho_\lambda'}\lambda \sqrt{r}}{4\Lambda_{\rm min}(\mathbf{H}(\bm{\mathcal{B}}^0))-3\mu},~~\|\wh{\bm{\mathcal{B}}}-\bm{\mathcal{B}}^0\|_*\le C\frac{C_{\rho_\lambda'}\lambda r}{4\Lambda_{\rm min}(\mathbf{H}(\bm{\mathcal{B}}^0))-3\mu},
$$
and the prediction error bound is
\begin{equation}\notag
\begin{aligned}
&{\rm PE}(\wh{\bm{\mathcal{B}}};\bm{\mathcal{B}}^0)\le C_{\rho_\lambda'}^2\lambda^2r\Big( \frac{c}{4\Lambda_{\rm min}(\mathbf{H}(\bm{\mathcal{B}}^0))-3\mu}+\frac{c\mu}{(4\Lambda_{\rm min}(\mathbf{H}(\bm{\mathcal{B}}^0))-3\mu)^2}\Big).
\end{aligned}
\end{equation}
\end{coro}
\begin{rmk}
The above results require that the cumulant function $\psi$ has bounded second derivative, $|\psi''(s)|\le c$ for any $s$, and its third order derivative decays sufficiently fast, $|\psi'''(s)|\le|s|^{-1}$ for $|s|> 1$, which are common for GLMs in many literature \citep{FAN2019177,JMLR:v16:loh15a}. Notably, these conditions hold for most members of the GLMs, such as linear regression, logistic regression, log-linear model and so on, but except for Poisson regression. 
\end{rmk}

Similarly, the t-TNN regularized estimator satisfies the following results. Notably, for the case of matrix, Proposition \ref{proGLMs} is consistent with results of Theorem 2 in \cite{FAN2019177}.

\begin{prp}\label{proGLMs}
Suppose that $\vx$ is sub-Gaussian, $0<\Lambda_{\rm min}(\mathbf{H}(\bm{\mathcal{B}}^0))$, $|\psi''(s)|\le c$ for any $s\in\mathbb{R}$ with a positive constant $c$ and $|\psi'''(s)|\le|s|^{-1}$ for $|s|> 1$. Let $\lambda\asymp \sqrt{dd_3/n}$, with high probability, we have
$$
\|\wh{\bm{\mathcal{B}}}-\bm{\mathcal{B}}^0\|_F\le C\frac{\lambda \sqrt{r}}{\Lambda_{\rm min}(\mathbf{H}(\bm{\mathcal{B}}^0))},~~\|\wh{\bm{\mathcal{B}}}-\bm{\mathcal{B}}^0\|_*\le C\frac{\lambda r}{\Lambda_{\rm min}(\mathbf{H}(\bm{\mathcal{B}}^0))},
$$
and
$$
{\rm PE}(\wh{\bm{\mathcal{B}}};\bm{\mathcal{B}}^0)\le C \frac{\lambda^2 r}{\Lambda_{\rm min}(\mathbf{H}(\bm{\mathcal{B}}^0))}.
$$
\end{prp}

\subsection{Adaptive huber tensor regression}\label{Sec-AHTR}
It is well-known that the above methods are sensitive to the outliers and heavy-tailed noise. To enhance robustness of estimator, Huber \citep{Huber73} proposed Huber loss and Sun et al. \citep{AHR19} further developed adaptive Huber regression where the robustification parameter is adapted by the the sample size, dimension and moments of noises. Therefore, we aim to extend the method of adaptive Huber to tensor regression. It is worth mentioning that it is nontrivial extension due to the complex structure within tensor.

The population and empirical risk function are respectively
$$
L(\bm{\mathcal{B}})=\mathbb{E}[l_{\upsilon}(y-\langle\bm{\mathcal{X}},\bm{\mathcal{B}}\rangle)],~~L_n(\bm{\mathcal{B}})=\frac{1}{n}\sum_{i=1}^n l_{\upsilon}(y_i-\langle\bm{\mathcal{X}}_i,\bm{\mathcal{B}}\rangle),
$$
where $l_{\upsilon}$ is the Huber loss,
\begin{equation}\notag
l_{\upsilon}(x)=
\begin{cases}
\frac{x^2}{2},&|x|\le\upsilon,\\
\upsilon|x|-\frac{\upsilon^2}{2},&|x|>\upsilon.
\end{cases}
\end{equation}
with robustification parameter $\upsilon>0$. It is easy to see that the smaller $\upsilon$ the more robustness but away from the mean estimation. Conversely, the robustness will decay as $\upsilon$ becomes large, and it reduces to the squared loss when $\upsilon$ goes to infinity with losing of robustness. Hence, it is crucial to set the value of $\upsilon$ via data-driven method. The proposed estimator is defined by
\begin{equation}\label{AHTR}
\wh{\bm{\mathcal{B}}}\in\mathop{\rm argmin} \frac{1}{n}\sum_{i=1}^n l_{\upsilon}(y_i-\langle\bm{\mathcal{X}}_i,\bm{\mathcal{B}}\rangle)+\rho_\lambda(\bm{\mathcal{B}}).
\end{equation}

Let $\mathbf{H}_n:=\frac{1}{n}\sum_{i=1}^n\vx_i\vx_i^{\mathsf T}$ and $\mathbf{H}:=\mathbb{E}[\mathbf{H}_n]$. Then, we have following results with some mild conditions.
\begin{coro}\label{AHTRB}
Suppose that $\vx$ is sub-Gaussian, $3\mu<\Lambda_{\rm min}(\mathbf{H})$, $\mathbb{E}[\epsilon|\bm{\mathcal{X}}]=0$ and $\mathbb{E}[|\epsilon|^{1+\delta}|\bm{\mathcal{X}}]\le\sigma_{\delta}$ for any $\delta\in(0,1]$. Denoting $c_\delta=(\sigma_\delta)^{1/(1+\delta)}$, taking $\lambda\asymp c_\delta(dd_3/n)^{\delta/(1+\delta)}$, $\upsilon\asymp c_\delta(n/dd_3)^{1/(1+\delta)}$  and $n\ge C\upsilon^2d^2d_3/\beta^2$, any stationary point $\wh{\bm{\mathcal{B}}}$ of program  \eqref{AHTR} satisfies
$$
\|\wh{\bm{\mathcal{B}}}-\bm{\mathcal{B}}^0\|_F\le C\frac{C_{\rho_\lambda'}\lambda \sqrt{r}}{\Lambda_{\rm min}(\mathbf{H})-3\mu},~~\|\wh{\bm{\mathcal{B}}}-\bm{\mathcal{B}}^0\|_*\le C\frac{C_{\rho_\lambda'}\lambda r}{\Lambda_{\rm min}(\mathbf{H})-3\mu},
$$
and
$$
{\rm PE}(\wh{\bm{\mathcal{B}}};\bm{\mathcal{B}}^0)\le C_{\rho_\lambda'}^2\lambda^2r\Big( \frac{c}{\Lambda_{\rm min}(\mathbf{H})-3\mu}+\frac{c\mu}{(\Lambda_{\rm min}(\mathbf{H})-3\mu)^2}\Big),
$$
with high probability.
\end{coro}
\begin{rmk}
The condition assumed on error term is relatively mild, which includes the conditional heteroscedastic models where we allow that $\epsilon$ and $\bm{\mathcal{X}}$ are not independent. Note that, when $\delta=1$, the Huber loss based estimator could achieve the statistical rate in Corollary \ref{LTRB} since the term $\lambda$ is of the same order. It is clear that the noises are more heavy-tailed $(0<\delta<1)$, the statistical rate of Huber estimator is slower. 
\end{rmk}

The t-TNN regularized Huber estimator satisfies the following bounds.
\begin{prp}\label{proHuber}
Suppose that $\vx$ is sub-Gaussian,$0<\Lambda_{\rm min}(\mathbf{H})$, $\mathbb{E}[\epsilon|\bm{\mathcal{X}}]=0$ and $\mathbb{E}[|\epsilon|^{1+\delta}|\bm{\mathcal{X}}]\le\sigma_{\delta}$ for any $\delta\in(0,1]$. Denoting $c_\delta=(\sigma_\delta)^{1/(1+\delta)}$, taking $\lambda\asymp c_\delta(dd_3/n)^{\delta/(1+\delta)}$, $\upsilon\asymp c_\delta(n/dd_3)^{1/(1+\delta)}$  and $n\ge C\upsilon^2d^2d_3/\beta^2$, we have
$$
\|\wh{\bm{\mathcal{B}}}-\bm{\mathcal{B}}^0\|_F\le C\frac{\lambda \sqrt{r}}{\Lambda_{\rm min}(\mathbf{H})},~~\|\wh{\bm{\mathcal{B}}}-\bm{\mathcal{B}}^0\|_*\le C\frac{\lambda r}{\Lambda_{\rm min}(\mathbf{H})},
$$
and
$$
{\rm PE}(\wh{\bm{\mathcal{B}}};\bm{\mathcal{B}}^0)\le C \frac{\lambda^2r}{\Lambda_{\rm min}(\mathbf{H})},
$$
with high probability.
\end{prp}
Wang et al. \citep{WangSS2025} has developed adaptive Huber matrix regression and their the statistical rate is as follows,
\begin{equation}\label{WangHuberMat}
\|\wh{\vB}-\vB^0\|\le C\lambda\sqrt{r},
\end{equation}
where $\lambda\asymp c_\delta(d/n)^{\delta/(1+\delta)}$. It is clear that the results of Proposition \ref{proHuber} reduces to \eqref{WangHuberMat} by adding some constants independent on $r,d,n$ to $C$ when $d_3=1$.

\subsection{Robust tensor regression via correntropy-induced loss}\label{Sec-CIRTR}
In the following, we propose a novel robust tensor regression model based on the correntropy-induced loss (C-loss) function. The correntropy measures general similarity between two random variables and captures the higher-order statistical relations which are undetectable for  conventional second-order statistics. The concept of correntropy has been widely used in signal processing and so on. Wang et al. \citep{WangESL} proposed a exponential squared loss which is actually a variant of C-loss and established the asymptotic normality of resulting estimator for the robust variable selection, and Feng et al. \citep{MCCRJMLR} developed learning theory analysis for the nonparametric regression via C-loss. The practical and theoretical success motivates us to develop the robust tensor regression with C-loss.

Formally speaking, the C-loss is defined as
$
l_\sigma(x):=\sigma^2\{1-{\rm exp}(-x^2/\sigma^2)\}
$ where $\sigma$ can be regarded as the robustification parameter. It is easy to see that the C-loss is nonconvex. Note that, the parameter controls the robustness of the loss. To visualize this point, we plot the geometry of it together with the squared loss and
the absolute loss for comparison.
\begin{figure}[htbp]
\centering
\includegraphics[scale=0.6]{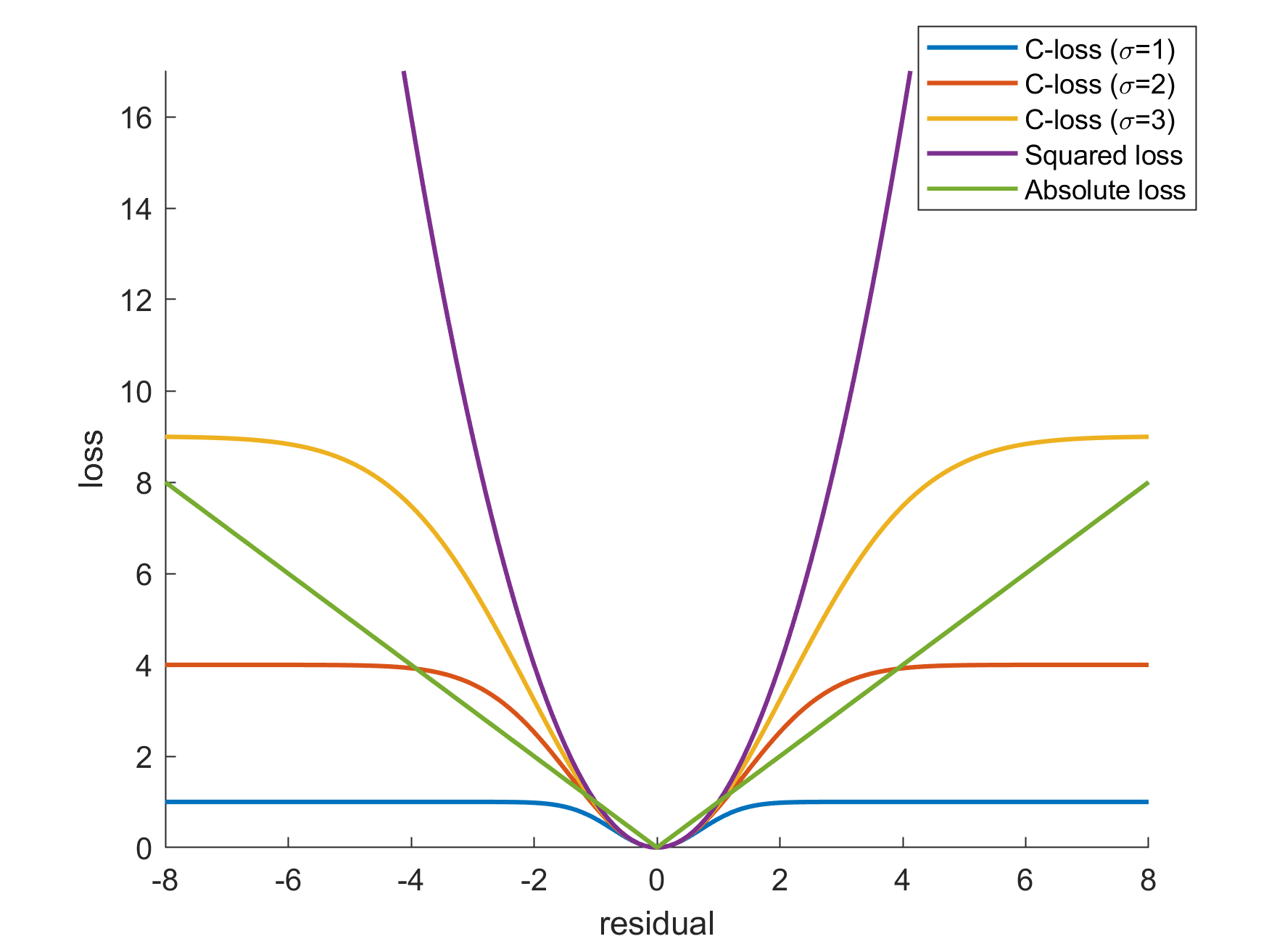}
\caption{Plots of the C-loss with different values of $\sigma$, the squared loss and the absolute loss.
}
\label{losscom}
\end{figure}
One can see that the C-loss works similarly as the squared loss for the small residuals, while it flattens when the residuals become large. The parameter $\sigma$ controls the geometry of the C-loss and thus determines the acceptable range for residuals. We will show that the parameter $\sigma$ also influence the statistical rate of the desired estimator, which implies that $\sigma$ plays a key balance between the robustness and the statistical efficiency.

The corresponding population and empirical risk function are given by
$$
L(\bm{\mathcal{B}})=\mathbb{E}[l_\sigma(y-\langle\bm{\mathcal{X}},\bm{\mathcal{B}}\rangle)],~~L_n(\bm{\mathcal{B}})=\frac{1}{n}\sum_{i=1}^nl_\sigma(y_i-\langle\bm{\mathcal{X}}_i,\bm{\mathcal{B}}\rangle).
$$
The nonconvex regularized estimator is defined by
\begin{equation}\label{CIRRTR}
\wh{\bm{\mathcal{B}}}\in\mathop{\rm argmin} \frac{1}{n}\sum_{i=1}^n l_{\sigma}(y_i-\langle\bm{\mathcal{X}}_i,\bm{\mathcal{B}}\rangle)+\rho_\lambda(\bm{\mathcal{B}}).
\end{equation}

\begin{coro}\label{CIRRTRB}
Suppose that $\vx$ and $\epsilon$ are sub-Gaussian with $M_X:={\rm sup}_{\|\bm{\mathcal{B}}\|_F\le 1}\|\langle \bm{\mathcal{X}},\bm{\mathcal{B}} \rangle\|_{\psi_2}<\infty$ for some positive constant $M_X$. Let $3\mu\le4\tau_1$, $n\ge Crdd_3/\beta^2$ and taking $\lambda\asymp\sqrt{dd_3/n}$, for any stationary point $\wh{\bm{\mathcal{B}}}$ of the nonconvex program \eqref{CIRRTR}, with high probability, we have
$$
\|\wh{\bm{\mathcal{B}}}-\bm{\mathcal{B}}^0\|_F\le C\frac{C_{\rho_\lambda'}\lambda \sqrt{r}}{4\tau_1-3\mu},~~\|\wh{\bm{\mathcal{B}}}-\bm{\mathcal{B}}^0\|_*\le C\frac{C_{\rho_\lambda'}\lambda r}{4\tau_1-3\mu},
$$
and
$$
{\rm PE}(\wh{\bm{\mathcal{B}}};\bm{\mathcal{B}}^0)\le C_{\rho_\lambda'}^2\lambda^2r\Big( \frac{c}{4\tau_1-3\mu}+\frac{c\mu}{(4\tau_1-3\mu)^2}\Big),
$$
where $\tau_1$ is a positive constant with respect to $M_X$, $\beta$, $\Lambda_{\rm min}(\mathbf{H})$ and $\sigma$. The explicit form of $\tau_1$ is given in the proof.
\end{coro}

\begin{rmk}
The estimation bound is inversely proportional to the curvature $\tau_1$, which implies that the higher curvature the faster statistical rate. With explicit form of $\tau_1$ in the proof, one can see that $\tau_1$ increases, but has a upper bound, as $\sigma$ becomes larger. It suggests that the larger $\sigma$ will incur the faster statistical rate, however, the robustness will reduces. Therefore, the parameter $\sigma$ balances the robustness of loss and the statistical rate of estimation. 
\end{rmk}

The same argument can be found for the t-TNN regularized estimator. If $d_3=1$, the results of Proposition \ref{prp:cie} degrades to that of \cite{SongRobustHT}.
\begin{prp}\label{prp:cie}
Suppose that $\vx$ and $\epsilon$ are sub-Gaussian with $M_X:={\rm sup}_{\|\bm{\mathcal{B}}\|_F\le 1}\|\langle \bm{\mathcal{X}},\bm{\mathcal{B}} \rangle\|_{\psi_2}<\infty$ for some positive constant $M_X$. Let  $n\ge Crdd_3/\beta^2$ and taking $\lambda\asymp\sqrt{dd_3/n}$, for any stationary point $\wh{\bm{\mathcal{B}}}$ of t-TNN regularized robust tensor regression problem, we have,
$$
\|\wh{\bm{\mathcal{B}}}-\bm{\mathcal{B}}^0\|_F\le C\frac{\lambda \sqrt{r}}{\tau_1},~~\|\wh{\bm{\mathcal{B}}}-\bm{\mathcal{B}}^0\|_*\le C\frac{\lambda r}{\tau_1},
$$
and
$$
{\rm PE}(\wh{\bm{\mathcal{B}}};\bm{\mathcal{B}}^0)\le  C\frac{\lambda^2r}{\tau_1},
$$
with high probability.
\end{prp}

\subsection{Robust logistic tensor regression via minimum distance criterion}\label{Sec-RLTR}
Finally, we consider the problem of robust logistic tensor regression. As Croux et al. \citep{CROUX2002377} and Feng et al. \citep{Feng2014RobustLR} claimed, if the collected samples are contaminated, the resulting estimator by MLE for logistic regression (LR) will have implosion breakdown. To enhance the robustness of LR, Chi et al. \citep{Chi2014} proposed a data-driven robust LR by the minimum distance criterion, which is actually a special case of density power divergence \citep{1998Robust}.

We first briefly review the minimum distance criterion and we refer readers to \citep{Scott2009L2E} for more details. Denote the estimated unknown true density as $f(x)$ and a member of family of density specified by a parameter $\theta$ as $\hat{f}(x|\theta)$. Rather than using MLE, seeking $\theta$ by minimizing integrated square error:
$$
\int[\hat{f}(x|\theta)-f(x)]^2dx.
$$
Expanding the above equation and ignoring some terms independent of $\theta$, we can derive the final estimation criterion in sample version:
$$
L_n(\theta)=\int\hat{f}^2(x|\theta)dx -\frac{2}{n}\sum_{i=1}^n\hat{f}(x_i|\theta).
$$
Thus, we call minimizing the above criterion as minimum distance criterion.

Now, we extend this concept to logistic tensor regression. Let $y$ be the binary response variable and $\bm{\mathcal{X}}$ be the tensor predictor. The conditional distribution of $y$, given $\bm{\mathcal{X}}$, is formulated as
\begin{equation}\notag
P(y|\bm{\mathcal{X}};\bm{\mathcal{B}})=
\begin{cases}
1-\sigma(\langle\bm{\mathcal{X}},\bm{\mathcal{B}}\rangle),& y=0,\\
\sigma(\langle\bm{\mathcal{X}},\bm{\mathcal{B}}\rangle),&y=1,
\end{cases}
\end{equation}
where $\sigma(x)=1/(1+e^{-x})$ for all $x\in\mathbb{R}$.
By the minimum distance criterion, the corresponding empirical risk function is given by
$$
L_n(\bm{\mathcal{B}})=\frac{1}{n}\sum_{i=1}^n\Big[\sum_{y\in\{0,1\}}P^2(y|\bm{\mathcal{X}}_i;\bm{\mathcal{B}})-2P(y_i|\bm{\mathcal{X}}_i;\bm{\mathcal{B}})  \Big].
$$
Furthermore, we can rewrite it in compact form by adding an additive and independent constant,
$$
L_n(\bm{\mathcal{B}})=\frac{1}{n}\sum_{i=1}^n(y_i-\sigma(\langle\bm{\mathcal{X}}_i,\bm{\mathcal{B}}\rangle))^2.
$$
One can see that $L_n$ is nonconvex. The population risk function is $L(\bm{\mathcal{B}})=\mathbb{E}[L_n(\bm{\mathcal{B}})]$, and the estimator is defined as
\begin{equation}\label{RLTR}
\wh{\bm{\mathcal{B}}}\in\mathop{\rm argmin} \frac{1}{n}\sum_{i=1}^n(y_i-\sigma(\langle\bm{\mathcal{X}}_i,\bm{\mathcal{B}}\rangle))^2+\rho_\lambda(\bm{\mathcal{B}}).
\end{equation}
\begin{coro}\label{RLTRB}
Suppose that $\vx$ is sub-Gaussian with $M_X:={\rm sup}_{\|\bm{\mathcal{B}}\|_F\le 1}\|\langle \bm{\mathcal{X}},\bm{\mathcal{B}} \rangle\|_{\psi_2}<\infty$ for some positive constant $M_X$. Taking $\lambda\asymp\sqrt{dd_3/n}$ and assuming $3\mu\le4\tau_1$, $n\ge Crdd_3/\beta^2$, for any stationary point $\wh{\bm{\mathcal{B}}}$ of the nonconvex problem \eqref{RLTR}, with high probability, we have
$$
\|\wh{\bm{\mathcal{B}}}-\bm{\mathcal{B}}^0\|_F\le C\frac{C_{\rho_\lambda'}\lambda \sqrt{r}}{4\tau_1-3\mu},~~\|\wh{\bm{\mathcal{B}}}-\bm{\mathcal{B}}^0\|_*\le C\frac{C_{\rho_\lambda'}\lambda r}{4\tau_1-3\mu},
$$
and
$$
{\rm PE}(\wh{\bm{\mathcal{B}}};\bm{\mathcal{B}}^0)\le C_{\rho_\lambda'}^2\lambda^2r\Big( \frac{c}{4\tau_1-3\mu}+\frac{c\mu}{(4\tau_1-3\mu)^2}\Big),
$$
where $\tau_1$ is a positive constant with respect to $M_X$, $\beta$ and $\Lambda_{\rm min}(\mathbf{H})$. The explicit form of $\tau_1$ is given in the proof.
\end{coro}
Similarly, for the t-TNN penalty based estimator, the following results holds.
\begin{prp}
Suppose that $\vx$ is sub-Gaussian with $M_X:={\rm sup}_{\|\bm{\mathcal{B}}\|_F\le 1}\|\langle \bm{\mathcal{X}},\bm{\mathcal{B}} \rangle\|_{\psi_2}<\infty$ for some positive constant $M_X$. Let  $n\ge Crdd_3/\beta^2$ and taking $\lambda\asymp\sqrt{dd_3/n}$, for any stationary point $\wh{\bm{\mathcal{B}}}$ of t-TNN regularized robust logistic tensor regression problem, we have,
$$
\|\wh{\bm{\mathcal{B}}}-\bm{\mathcal{B}}^0\|_F\le C\frac{\lambda \sqrt{r}}{\tau_1},~~\|\wh{\bm{\mathcal{B}}}-\bm{\mathcal{B}}^0\|_*\le C\frac{\lambda r}{\tau_1},
$$
and
$$
{\rm PE}(\wh{\bm{\mathcal{B}}};\bm{\mathcal{B}}^0)\le  C\frac{\lambda^2r}{\tau_1},
$$
with high probability.
\end{prp}

\section{Numerical studies}\label{Sec5}
In this section, we report the results of experiments on the synthetic and real-world data to demonstrate the effectiveness of our proposed models. In regression paradigm, we compare the square loss, Huber loss and C-loss based regression methods (Section \ref{Sec-LTR}, \ref{Sec-AHTR} and \ref{Sec-CIRTR}) under the proposed nonconvex penalty function, with t-TNN regularized counterparts. In classification task, the proposed two logistic regression models (Section \ref{Sec-GLM} and \ref{Sec-RLTR}) are compared with t-TNN based counterparts. For ease of illustration, we indicate the least square regression with LSR, the adaptive Huber regression with AHR, the regression based correntropy-induced loss with CIR, the naive logistic regression with LR and the robust logistic regression via minimum distance criterion with RLR. For the regression models, we take the ordinary least square solution as the initial value of $\bm{\mathcal{B}}$, and set $\bm{\mathcal{B}}_0={\bm 0}$ in the classification task.
For each simulation, we report the average results of 100 independent replications, and all tuning parameters are selected by using $5$-fold cross-validation.

\subsection{Simulation}

\noindent{\textbf{(1) Linear regression:}} In this study, the data $\{(y_i,\bm{\mathcal{X}}_i)\}_{i=1}^n$ are generated from the below model,
$$
y_i=\langle\bm{\mathcal{X}}_i,\bm{\mathcal{B}}^0\rangle+\epsilon_i,~~i=1,\cdots,n.
$$
The $r$-tubal-rank true tensor coefficient $\bm{\mathcal{B}}^0=\bm{\mathcal{C}}_1*\bm{\mathcal{C}}_2$ with $\bm{\mathcal{C}}_1\in\mathbb{R}^{d_1\times r\times d_3}$ and $\bm{\mathcal{C}}_2\in\mathbb{R}^{r\times d_2\times d_3}$, where each element is independently sampled from the standard Gaussian distribution. In addition, the entries of $\bm{\mathcal{X}}_i$ also follow $N(0,1)$, and the mechanism of noise will be further specified in the following studies.

\textbf{(a) Effect of nonconvex penalty:} In this example, we aim to test the performance of various regularization terms. We consider the following noise distributions: (\rmnum{1}) light-tailed: the standard noise assumption, $N(0,1)$; (\rmnum{2}) heavy-tailed and symmetric: the $t$-distribution with 3 degrees of freedom, $t(3)$ and (\rmnum{3}) heavy-tailed and asymmetric: the Pareto distribution with scale parameter 3 and shape parameter 2, $Par(3,2)$. We set the sample size $n=2000$, dimension $d_1=d_2=20,d_3=3$ and tubal-rank $r=2$. We use the estimation error (Err) $\|\bm{\mathcal{B}}^0-\wh{\bm{\mathcal{B}}}\|_F$ and estimated rank $\hat{r}$ to quantify the performance of each model. Table \ref{Simu-T1} reports the averaged Err and $\hat{r}$.

\begin{table}[hptb]
\begin{center}
\caption{The averaged Err and $\hat{r}$ under different nonconvex penalties and various noise.}\label{Simu-T1}
\resizebox{\textwidth}{!}{
\begin{tabular}{cccccccccccccccc}\hline\hline
\multirow{2}*{Noise}&\multirow{2}*{Model} &\multicolumn{2}{c}{t-TNN} & \multicolumn{2}{c}{Geman} & \multicolumn{2}{c}{SCAD} & \multicolumn{2}{c}{Laplace} & \multicolumn{2}{c}{MCP}& \multicolumn{2}{c}{ETP}& \multicolumn{2}{c}{Logarithm}\\ \cline{3-16}
 & & Err & $\hat{r}$&  Err & $\hat{r}$&  Err & $\hat{r}$& Err & $\hat{r}$& Err & $\hat{r}$&Err & $\hat{r}$&Err & $\hat{r}$\\\hline
\multirow{3}*{$N(0,1)$}&LSR&	0.6107	&2.00&0.3567&	2.00&	0.3599&	2.00&	0.3588&	2.00&	0.3582&	2.00&	0.3580&	2.00&	0.3591&	2.00\\
&AHR&	0.6466&	2.00&0.3739&	2.00&	0.3843&	2.00&0.3771&	2.00	&0.3759&	2.00&	0.3748&	2.00&	0.3767&	2.00\\
 &CIR&	0.6367&	2.00&0.3705&	2.00&	0.3772&	2.00&	0.3737&	2.00&	0.3709&	2.00&	0.3671	&2.00&	0.3687&	2.00
\\
 \hline
 \multirow{3}*{$t(3)$}&LSR&	1.4518&	2.05&0.6174&	2.01&	0.6241&	2.00&	0.6083&	2.00&	0.6912&	2.00&	0.6297&	2.00&	0.6055&	2.00
\\
&AHR&	1.0742&	2.06&0.4580&	2.00&	0.4627&	2.00&	0.4557&	2.00&	0.4679&	1.98&	0.4583&	2.00&	0.4636&	2.00
\\
 &CIR&	0.9367&	2.08&0.4577&	2.00&	0.4635&	2.00&	0.4507&	2.00&	0.4564&	2.00&	0.4565&	2.00&	0.4604&	2.00
\\
 \hline
 \multirow{3}*{$Par(3,2)$}&LSR&	5.7846&	2.16&3.1703&	2.06&	3.1797&	2.00&	3.2584&	2.09&	3.1046&	2.05&	3.2466&	2.00&	3.2708&	2.00
\\
&AHR&	4.3929&	2.10&2.4233&	2.00&	2.3456&	2.00&	2.4713&	2.00&	2.2470&	2.00&	2.2388&	2.00&	2.2534&	2.08
\\
 &CIR&	4.2542&	2.09&2.3891&	2.00&	2.2736&	2.00&	2.3869&	2.02&	2.2176&	2.00&	2.2430&	2.00&	2.2447&	2.07
\\
\hline\hline
\end{tabular}
}
\end{center}
\end{table}

From the presented results in the above tables, when the noise follows standard Gaussian distribution, the LSR achieves the best performance. As the noise grows complex, AHR and CIR performs better than LSR, and in most cases CIR outperforms AHR. Meanwhile, the estimated rank of nonconvex penalty approaches to the true rank better, and the nonconvex penalty obtains the lower estimation error,  in comparison with t-TNN.  Note that there is little difference between various nonconvex regularizers in terms of estimation error and estimated rank. Hence, in the following studies, we only report the results yielded by MCP, and we further denote nonconvex penalty regularized LSR as NCRLSR. Similarly, we call NCRAHR for adaptive huber regression, NCRCIR for regression based on correntropy-induced loss, and TNNLSR, TNNAHR, TNNCIR for t-TNN based counterparts. Moreover, we plot the value of objective function in each iteration to visualize the convergence behavior of our proposed estimation algorithm. The convergence curve in Figure \ref{Simu-F1} is recorded from nonconvex penalty based regression models under standard Gaussian noise, together with t-TNN counterparts. As Figure \ref{Simu-F1} illustrated, the value of objective function decreases monotonically, which verify our convergence results in Section \ref{Sec3}.

\begin{figure}[!htb]
\centering
\subfigure[Convergence curve of the regression models with nonconvex regularizer]{
		\includegraphics[scale=0.8]{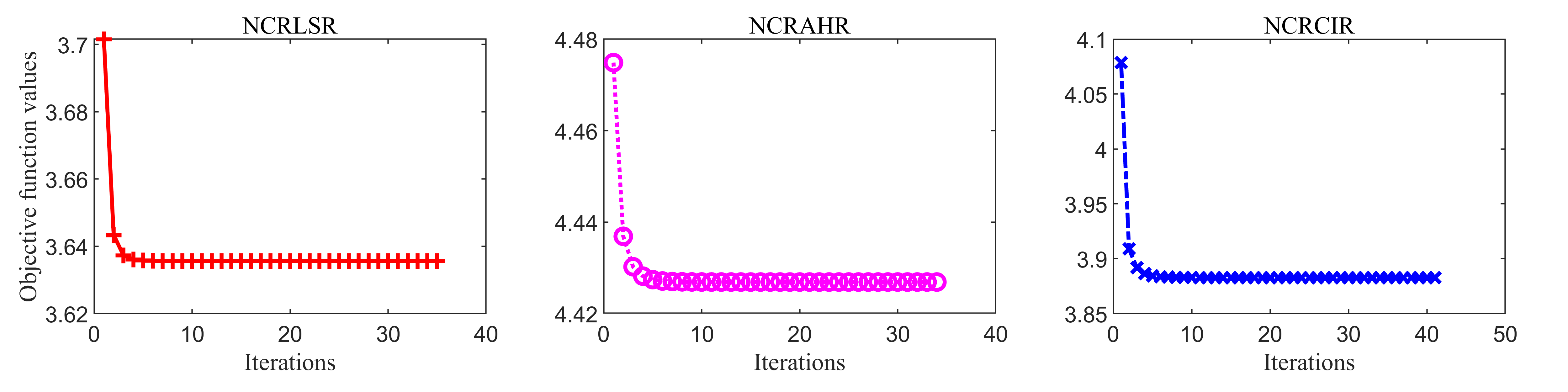}}\\
\subfigure[Convergence curve of the regression models with t-TNN penalty]{
		\includegraphics[scale=0.8]{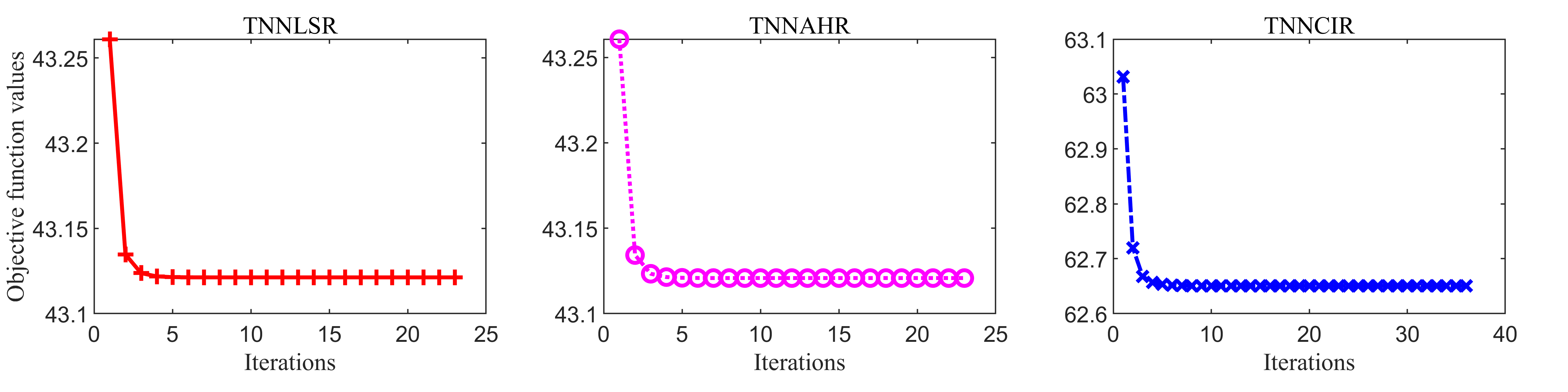}}
\caption{Convergence behavior of linear regression models.}
\label{Simu-F1}
\end{figure}

\textbf{(b) Effect of robustness:} Furthermore, we consider the conditional heteroscedastic model,
$$
y_i=\langle\bm{\mathcal{X}}_i,\bm{\mathcal{B}}^0\rangle+(1+|\bm{\mathcal{X}}_i(1,1,1)|)\epsilon_i,~~i=1,\cdots,n,
$$
where the noise still follows the above distributions. The sample size, dimension and tubal-rank is taken same value as previous example. From Table \ref{Simu-T2}, under the conditional heteroscedastic model, we can find that the performance of each model decays as noise becomes complex. Especially  for the Pareto distribution noise, the square loss based regression method breaks down. And in such case, TNNAHR performs better than TNNCIR ableit NCRCIR outperforms NCRAHR.
The nonconvex regulariers penalty could yields lower Err and good $\hat{r}$ for any noise. Hence, the nonconvex penalty not only yield accurate rank estimation but also enhance the capability of estimation, which implies it actually reduce the bias incurred by regularization.
\begin{table}[hptb]
\begin{center}
\caption{The averaged Err and $\hat{r}$ under conditional heteroscedastic model with different noise.}\label{Simu-T2}
\resizebox{\textwidth}{!}{
\begin{tabular}{ccccccccccccc}\hline\hline
\multirow{2}*{Noise}&\multicolumn{2}{c}{TNNLSR	} & \multicolumn{2}{c}{TNNAHR} & \multicolumn{2}{c}{TNNCIR} & \multicolumn{2}{c}{NCRLSR} & \multicolumn{2}{c}{NCRAHR}& \multicolumn{2}{c}{NCRCIR}\\ \cline{2-13}
  & Err & $\hat{r}$& Err & $\hat{r}$& Err & $\hat{r}$& Err & $\hat{r}$& Err & $\hat{r}$& Err & $\hat{r}$\\\hline
$N(0,1)$&1.0988& 2.05&	0.7640&	2.02&	0.5771&	2.01&	0.3598&	2.00&	0.2801	&2.00&	0.2396&	2.00\\
$t(3)$&1.3413&	2.10&	0.8584&	2.06&	0.6862&	2.09&	0.6007&	2.02&	0.3263&	2.00&	0.2677&	2.00\\
$Par(3,2)$&7.2703&	2.60&	4.6115&	3.00&	4.6190&	3.05&	4.1573&	2.24&	2.8604&	2.12&	2.8593&	2.10\\
\hline\hline
\end{tabular}
}
\end{center}
\end{table}

\textbf{(c) Effect of sample size:} In this experiment, we still consider the conditional heteroscedastic model but only using standard Gaussian noise. The effect of sample size on the estimation error is of interest for us. We vary $n$ from $2000,4000$ to $6000$, the dimension and tubal-rank is taken as above. Table \ref{Simu-T3} records the estimation error and estimated rank, one can see that the Err deduces as $n$ increases which verify our statistical rate, and the rank estimated by nonconvex penalty is more closer to the true rank than t-TNN.
\begin{table}[hptb]
\begin{center}
\caption{The averaged Err and $\hat{r}$ under conditional heteroscedastic model of standard Gaussian noise with different sample size.}\label{Simu-T3}
\footnotesize
\resizebox{\textwidth}{!}{
\begin{tabular}{ccccccccccccc}\hline\hline
\multirow{2}*{$n$}&\multicolumn{2}{c}{TNNLSR	} & \multicolumn{2}{c}{TNNAHR} & \multicolumn{2}{c}{TNNCIR} & \multicolumn{2}{c}{NCRLSR} & \multicolumn{2}{c}{NCRAHR}& \multicolumn{2}{c}{NCRCIR}\\ \cline{2-13}
  & Err & $\hat{r}$& Err & $\hat{r}$& Err & $\hat{r}$& Err & $\hat{r}$& Err & $\hat{r}$& Err & $\hat{r}$\\\hline
2000&1.0988&	2.05&	0.7640&	2.02&	0.5771&	2.01&	0.3598&	2.00&	0.2801&	2.00&	0.2396&	2.00\\
4000&0.5487&	2.06&	0.4164&	2.01&	0.3012&	2.01&	0.2464&	2.00&	0.1876&	2.00&	0.1534&	2.00\\
6000&0.3811&	2.02&	0.3291&	2.02&	0.2261&	2.03&	0.1781&	2.00&	0.1498&	2.00&	0.1209&	2.00\\
\hline\hline
\end{tabular}
}
\end{center}
\end{table}

\textbf{(d) Effect of tubal rank:} At last, we study the performance of various estimators if the true tubal rank is different. The true tubal rank is taken from $2,4$ to $8$, the dimension is set as above and the sample size takes $2000$. The conditional heteroscedastic model with standard Gaussian noise is still considered. From Table \ref{Simu-T4}, one can see that the estimation error increases as the tubal rank grows, which is consistent with our statistical results.
\begin{table}[hptb]
\begin{center}
\caption{The averaged Err and $\hat{r}$ under conditional heteroscedastic model of standard Gaussian noise with different tubal rank.}\label{Simu-T4}
\resizebox{\textwidth}{!}{
\begin{tabular}{ccccccccccccc}\hline\hline
\multirow{2}*{$r$}&\multicolumn{2}{c}{TNNLSR	} & \multicolumn{2}{c}{TNNAHR} & \multicolumn{2}{c}{TNNCIR} & \multicolumn{2}{c}{NCRLSR} & \multicolumn{2}{c}{NCRAHR}& \multicolumn{2}{c}{NCRCIR}\\ \cline{2-13}
  & Err & $\hat{r}$& Err & $\hat{r}$& Err & $\hat{r}$& Err & $\hat{r}$& Err & $\hat{r}$& Err & $\hat{r}$\\\hline
2&	0.9088&	2.05&	0.7640&	2.02&	0.5771&	2.01&	0.3598&	2.00&	0.2801&	2.00&	0.2396&	2.00\\
4&	1.0275&	4.06&	0.9652&	4.04&	0.7893&	4.08&	0.5302&	4.00&	0.4321&	4.00&	0.4179&	4.00\\
8&  1.2052&	8.05&	1.1174&	8.05&	1.0415&	8.02&	0.7875&	8.01&	0.7106&	8.00&	0.6719&	8.00\\
\hline\hline
\end{tabular}
}
\end{center}
\end{table}

To visualize the result of rank estimation, we adopt some colorful geometric shapes to serve as the ground-truth tensor parameter, including Cross, Square and T. We take the blue part in the shape as the true signal and others are nuisance, see Figure \ref{Simu-F2}. We consider the Gaussian distribution for Cross shape, the $t$ distribution for Square shape and the Pareto distribution for T shape. The dimension and sample size is taken as $d=20$ and $n=2000$, respectively. The image of true and estimated tensor parameter are demonstrated in Figure \ref{Simu-F2}, and the averaged Err are recorded in Table \ref{Simu-T5}. It is clear that the signal estimated by NCRCIR is closer to true signal than others. When the noise is complex, the least square estimation falls, and NCRLSR still outperforms TNNLSR. Table \ref{Simu-T5} also shows that our proposed NCRCIR achieves the best performance and the nonconvex penalty performs better than t-TNN.
\begin{figure}[htbp]\centering
\subfigure[\scriptsize True]{
\begin{minipage}[t]{0.1\linewidth}
			\centering
	\includegraphics[width=0.5in]{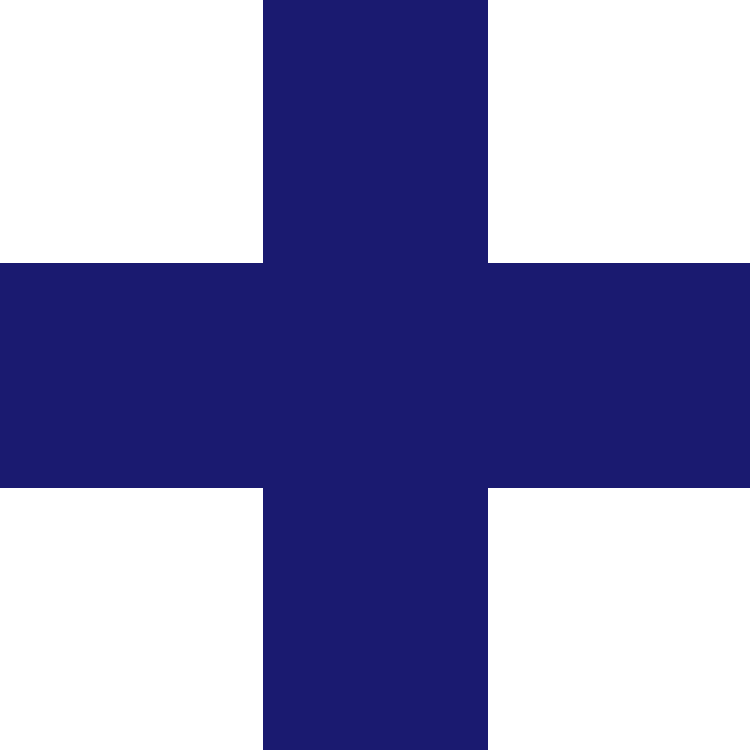}\\
	\includegraphics[width=0.5in]{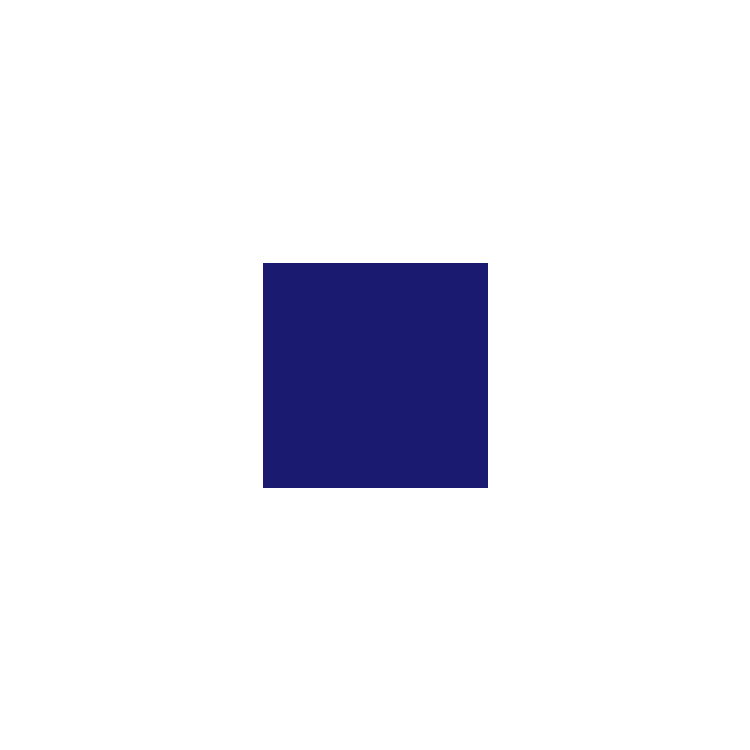}\\
	\includegraphics[width=0.5in]{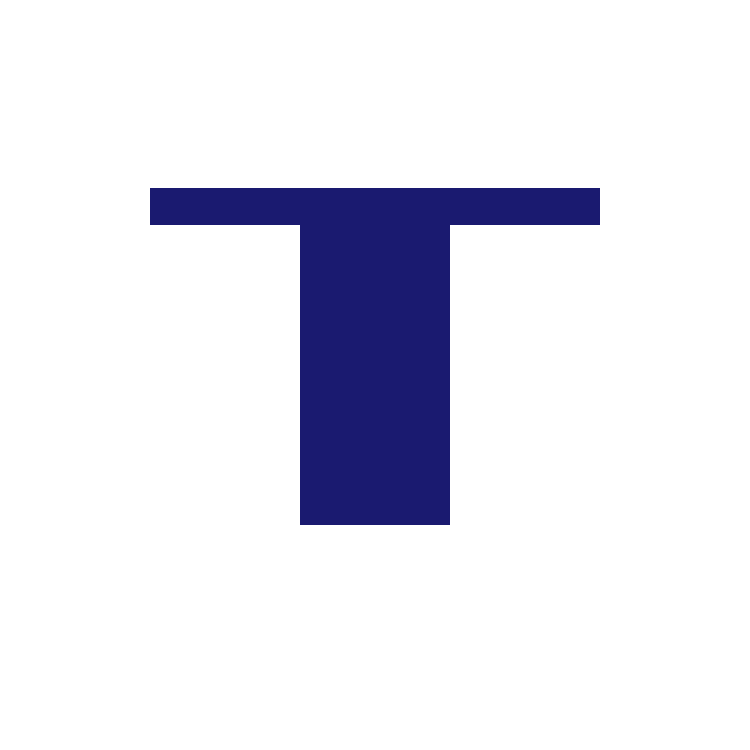}\\
\end{minipage}%
	}%
\hspace{4.6mm}
\centering
\subfigure[\scriptsize TNNLSR]{
		\begin{minipage}[t]{0.1\linewidth}
			\centering
	\includegraphics[width=0.5in]{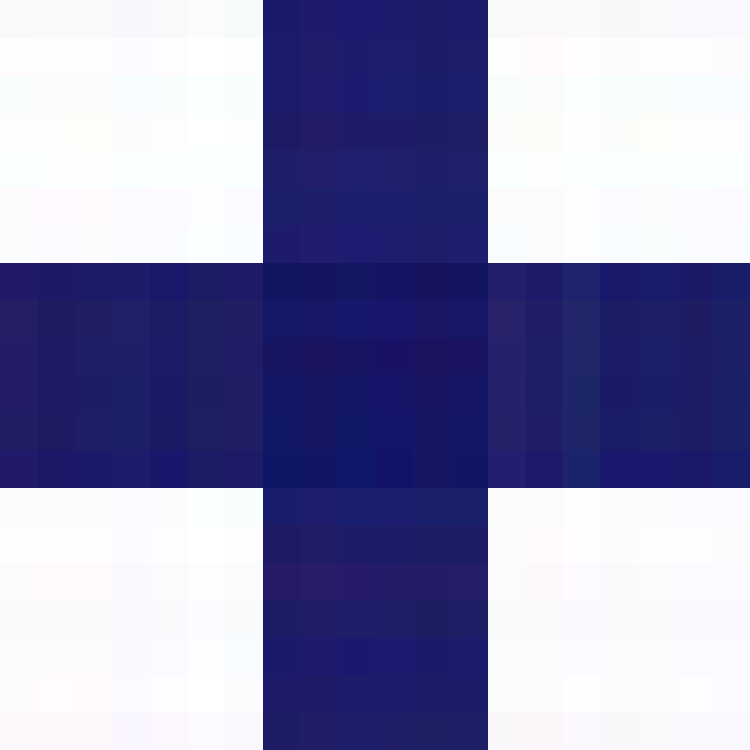}\\
	\includegraphics[width=0.5in]{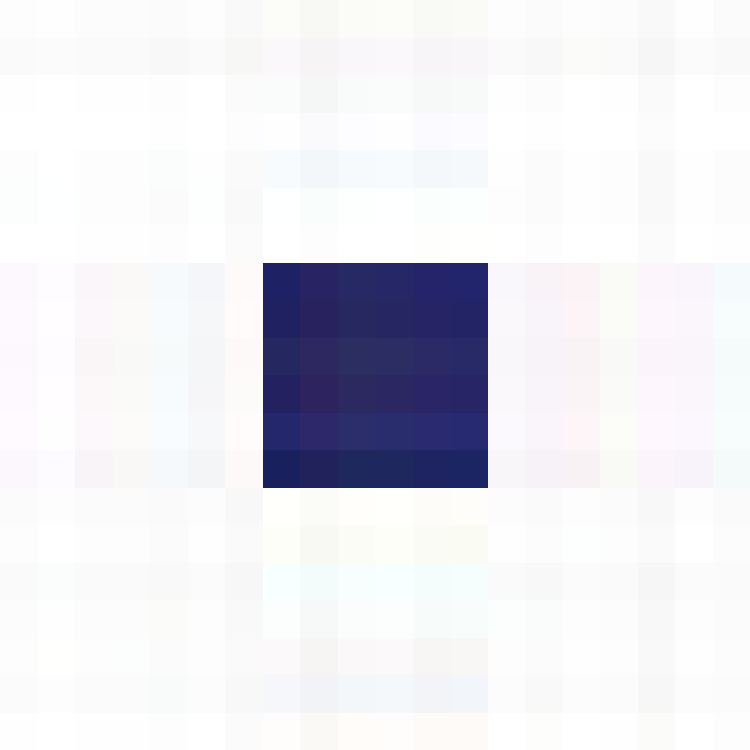}\\
	\includegraphics[width=0.5in]{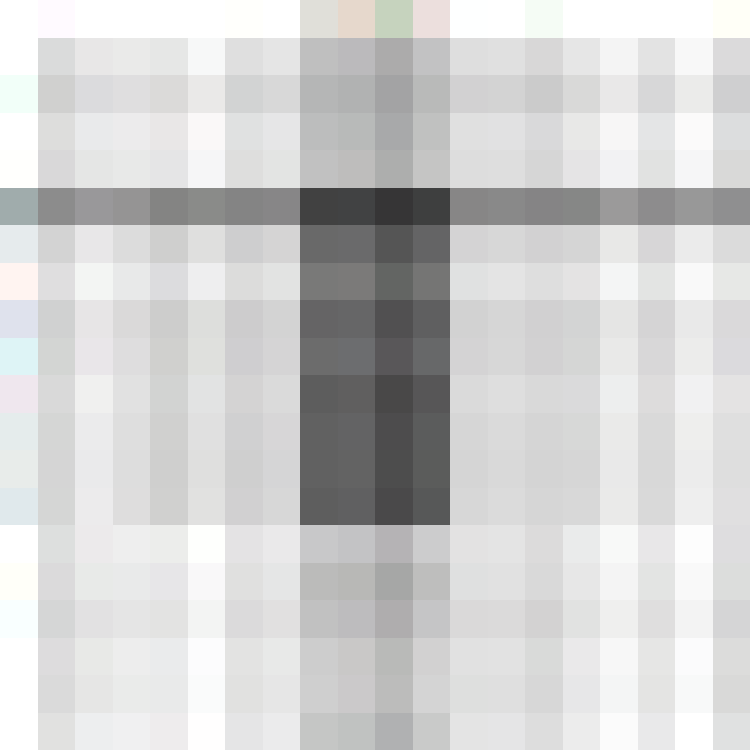}\\
		\end{minipage}%
	}%
\hspace{4.6mm}
\centering
\subfigure[\scriptsize TNNAHR]{
		\begin{minipage}[t]{0.1\linewidth}
			\centering
	\includegraphics[width=0.5in]{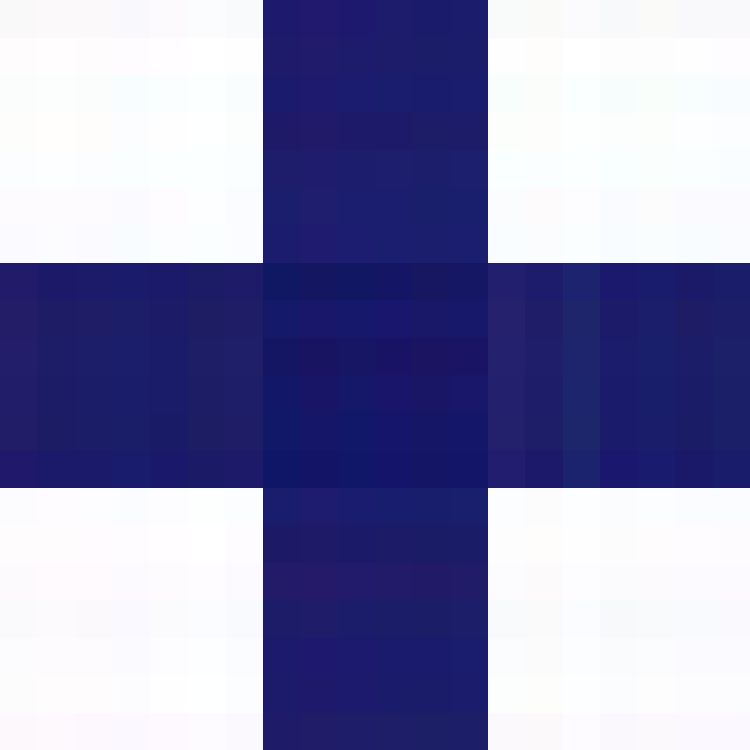}\\
	\includegraphics[width=0.5in]{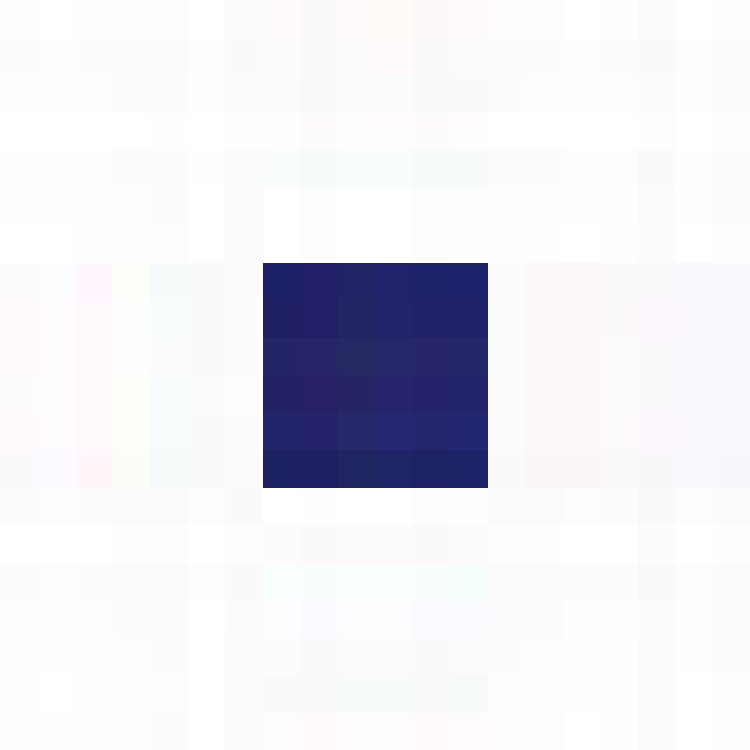}\\
	\includegraphics[width=0.5in]{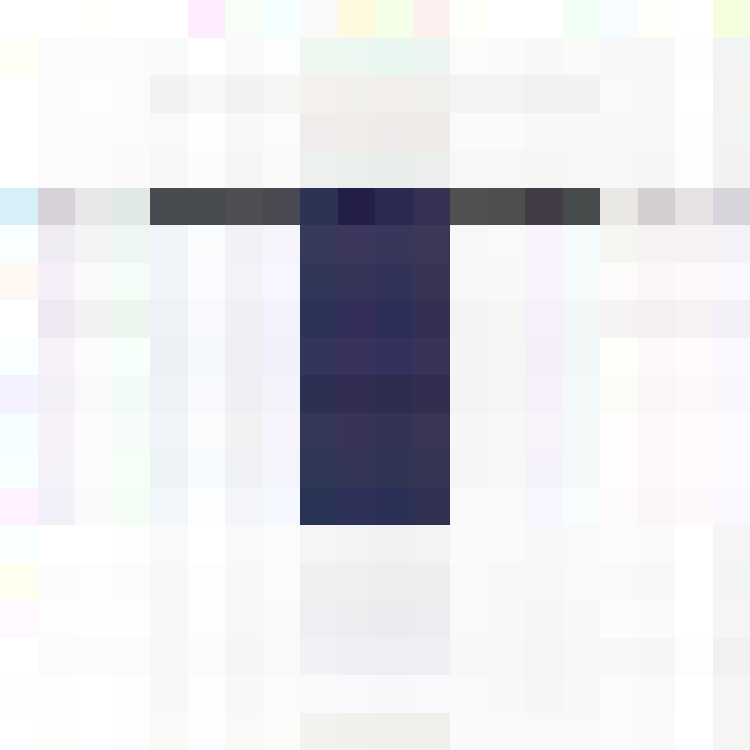}\\
		\end{minipage}%
	}%
\hspace{4.6mm}
\subfigure[\scriptsize TNNCIR]{
		\begin{minipage}[t]{0.1\linewidth}
			\centering
	\includegraphics[width=0.5in]{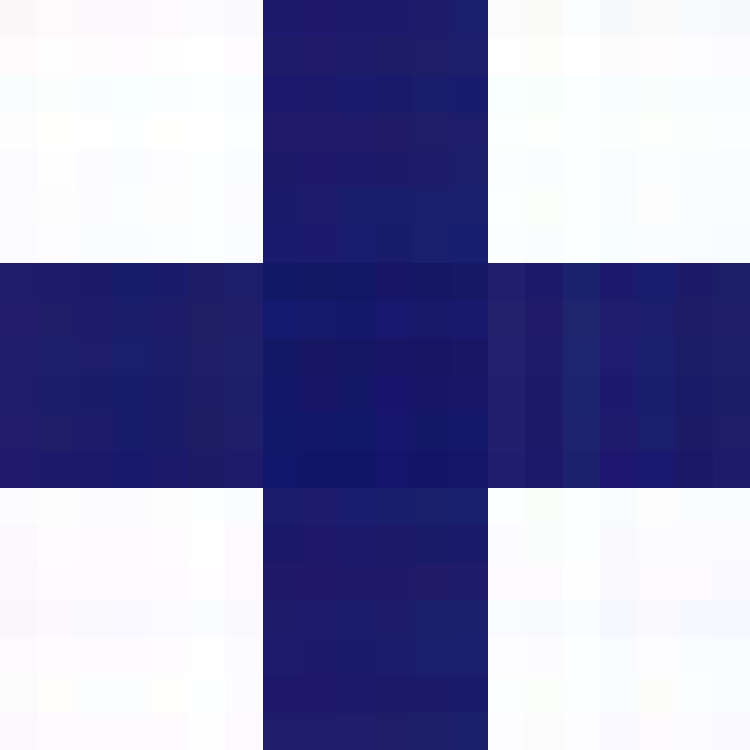}\\
	\includegraphics[width=0.5in]{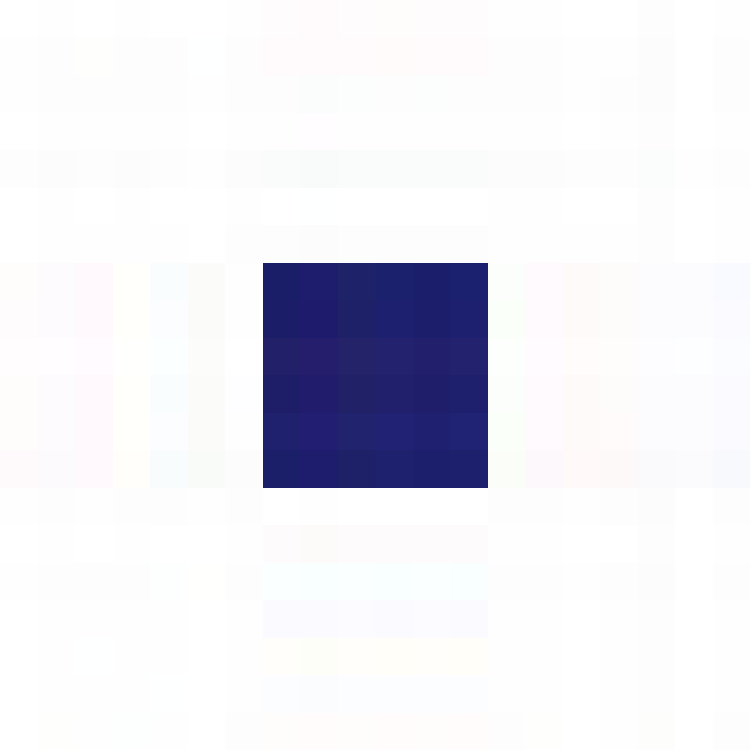}\\
	\includegraphics[width=0.5in]{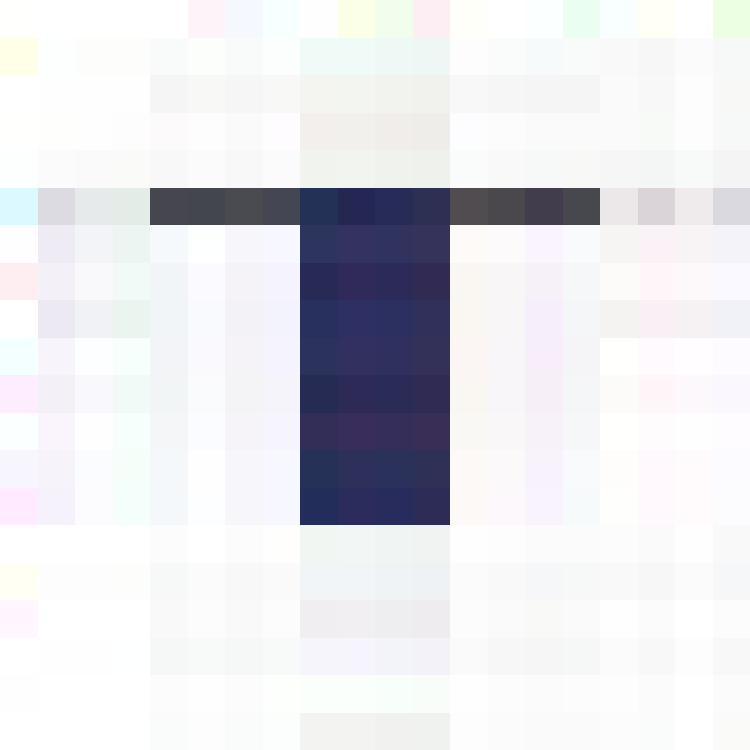}\\

		\end{minipage}%
	}%
\hspace{4.6mm}
\subfigure[\scriptsize NCRLSR]{
		\begin{minipage}[t]{0.1\linewidth}
			\centering
	\includegraphics[width=0.5in]{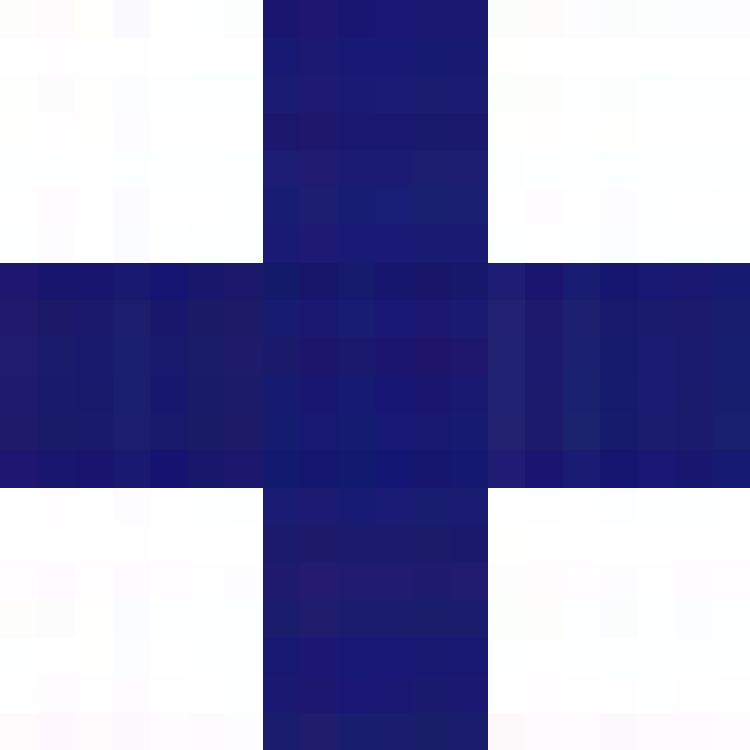}\\
	\includegraphics[width=0.5in]{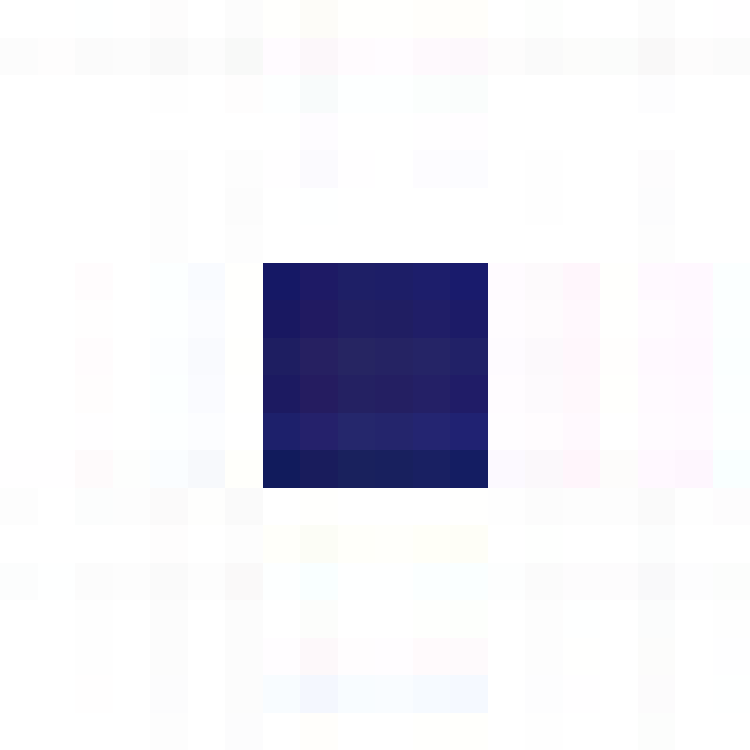}\\
	\includegraphics[width=0.5in]{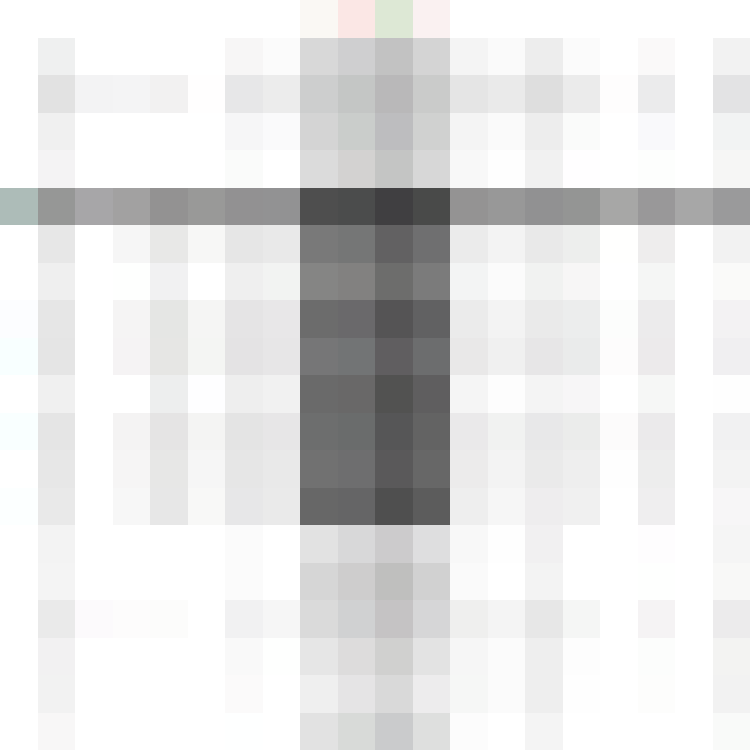}\\
		\end{minipage}%
	}%
\hspace{4.6mm}
\subfigure[\scriptsize NCRAHR]{
		\begin{minipage}[t]{0.1\linewidth}
			\centering
	\includegraphics[width=0.5in]{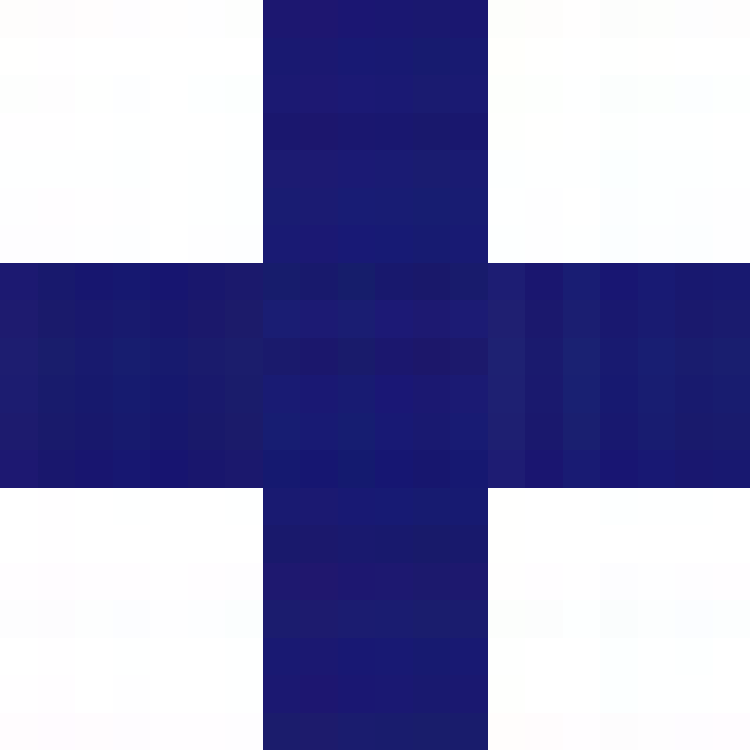}\\
	\includegraphics[width=0.5in]{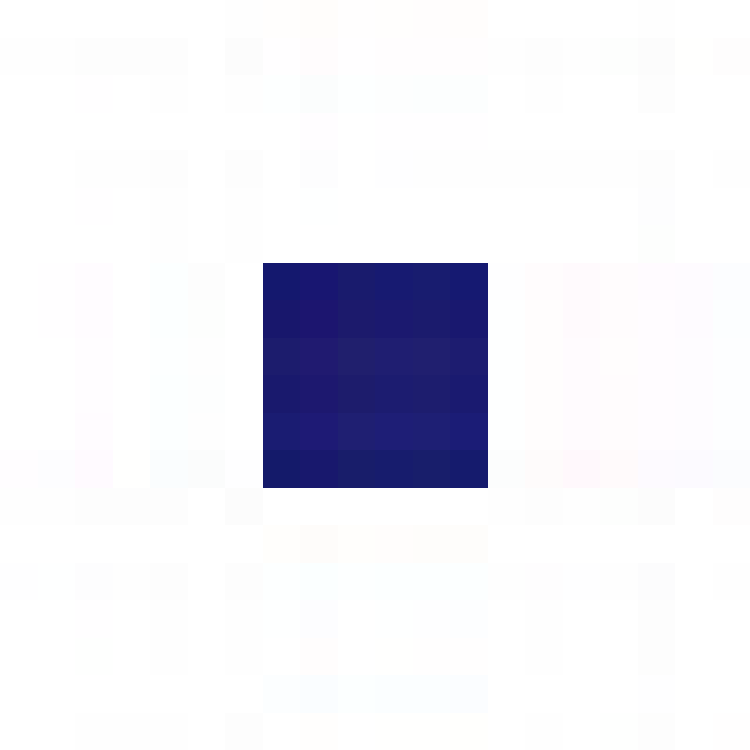}\\
	\includegraphics[width=0.5in]{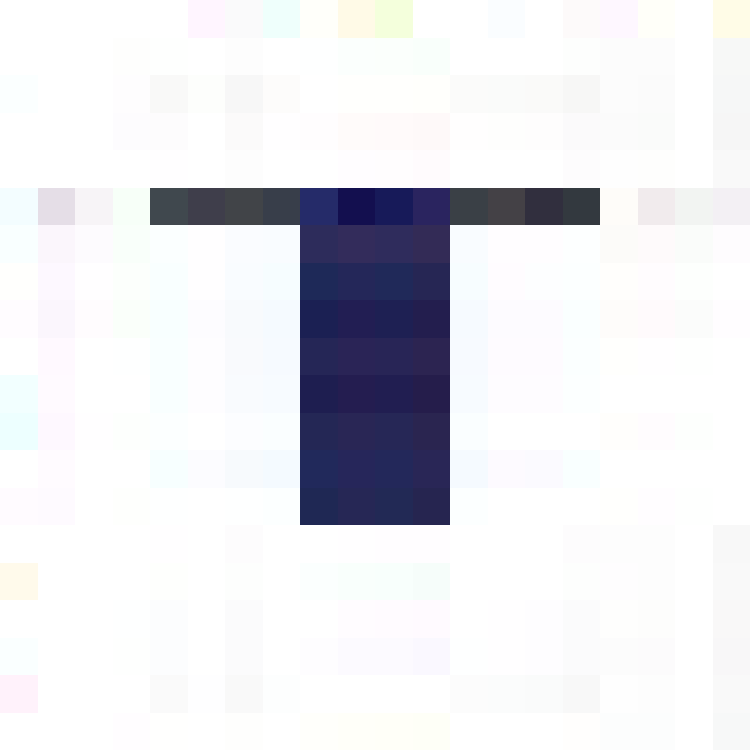}\\
		\end{minipage}%
	}%
\hspace{4.6mm}
\subfigure[\scriptsize NCRCIR]{
		\begin{minipage}[t]{0.1\linewidth}
			\centering
	\includegraphics[width=0.5in]{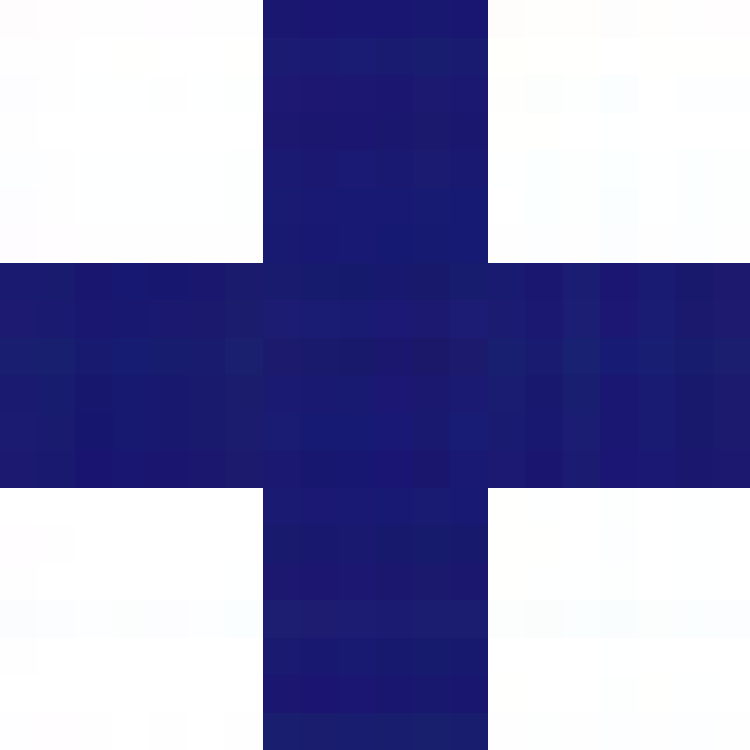}\\
	\includegraphics[width=0.5in]{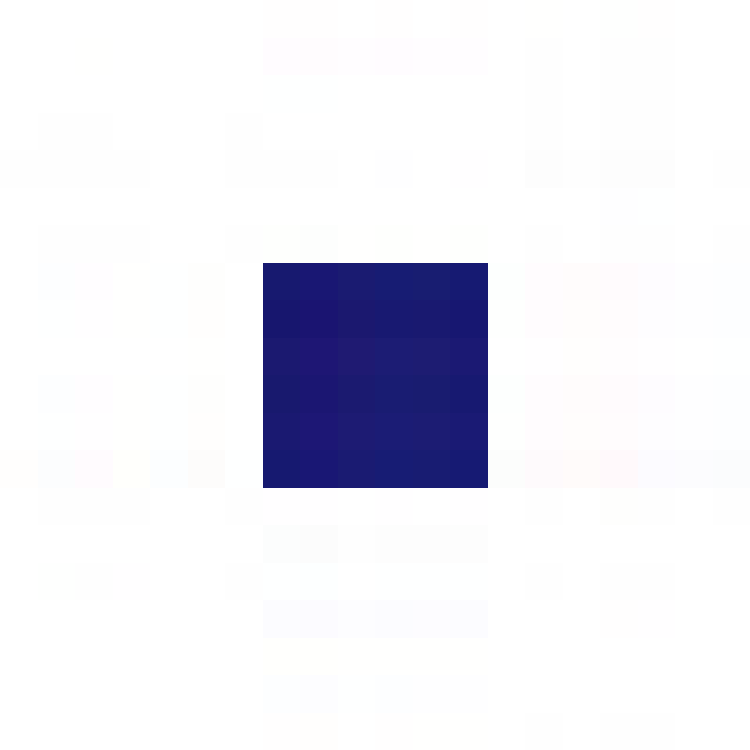}\\
	\includegraphics[width=0.5in]{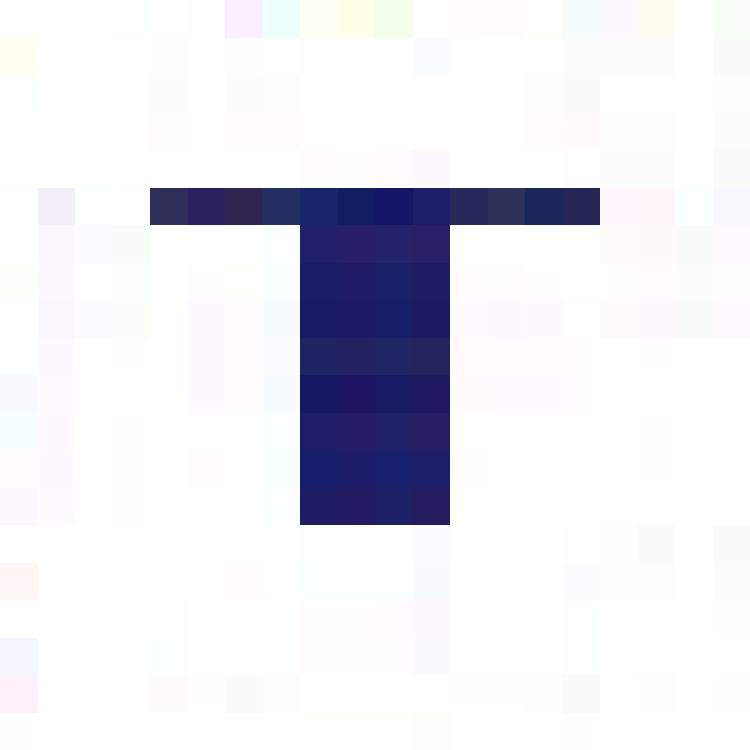}\\
		\end{minipage}%
	}%
\hspace{4.6mm}
\caption{The true signal and estimated signal of different methods. From top to bottom, the considered noise is from, in turn, the standard Gaussian, the student's t and the Pareto distribution.}
\label{Simu-F2}
\end{figure}

\begin{table}[hptb]
\begin{center}
\caption{The averaged Err under conditional heteroscedastic model with different noise distributions and geometric shapes.}\label{Simu-T5}
\footnotesize
\resizebox{\textwidth}{!}{
\begin{tabular}{cccccccc}\hline\hline
Geometric shape
&Noise & TNNLSR & TNNAHR & TNNCIR & NCRLSR& NCRAHR& NCRCIR\\ \hline
Cross&	$N(0,1)$&0.9111&	0.7029&	0.5782&	0.3836&	0.2844&	0.2342\\
Square&	$t(3)$	&1.2896&	0.7955&	0.6378&	0.5918&	0.3672&	0.2735\\
T&	$Par(3,2)$	&6.8923&	3.7759&	3.4930&	4.7819&	2.3495&	2.2749\\
\hline\hline
\end{tabular}
}
\end{center}
\end{table}

\noindent{\textbf{(2) Logistic regression:}} The synthetic data are generated by the following rule. The true tensor parameter $\bm{\mathcal{B}}^0$ is set as $\bm{\mathcal{B}}^0=\bm{\mathcal{C}}_1*\bm{\mathcal{C}}_2$ with $\bm{\mathcal{C}}_1\in\mathbb{R}^{d_1\times r\times d_3}$ and $\bm{\mathcal{C}}_2\in\mathbb{R}^{r\times d_2\times d_3}$ and we resize all the nonzero singular values of $\bm{\mathcal{B}}^0$ to $1$, where $r$ denotes the true tubal rank. The response $y_i$ follows the binary Bernoulli distribution, that is, $y_i\sim {\rm Bin}(1,{\rm exp}(\langle \bm{\mathcal{X}}_i ,\bm{\mathcal{B}}^0 \rangle)/(1+{\rm exp}(\langle \bm{\mathcal{X}}_i , \bm{\mathcal{B}}^0 \rangle)))$ where the elements of $\bm{\mathcal{X}}_i$ are i.i.d. samples of $N(0,1)$. 

\textbf{(a) Effect of sample size and tubal rank:} In this example, the sample size $n$ is varied from $2000,4000$ to $6000$, and the tubal rank $r$ from $2,4$ to $8$ with fixed dimension $d_1=d_2=20,d_3=3$. In addition, we randomly change $5\%$ samples in the class 0 to class 1, which means that there exist the issue of misspecification. The logarithmic estimation error (logErr) ${\rm log}\|\bm{\mathcal{B}}^0-\wh{\bm{\mathcal{B}}}\|_F$, prediction accuracy (Acc, $\%$) and estimated rank ($\hat{r}$) are used to evaluate the performance of various estimators. From Table \ref{Simu-T6}, as we  expected, the estimation error increases as the tubal rank grows, and decreases as the sample size becomes large. The NCRRLR outperforms others from respect of not only estimation error but also prediction accuracy. The convergence curve is given in Figure \ref{Simu-F3}.  One can see that the objective function value of proposed robust logistic regression reduces monotonically albeit it converges slowly compared with MLE since it is a nonconvex loss.

\begin{table}[hptb]
\begin{center}
\caption{The averaged logErr, Acc and $\hat{r}$ of different estimators.}\label{Simu-T6}
\footnotesize
\resizebox{\textwidth}{!}{
\begin{tabular}{cccccccccccccc}\hline\hline
\multirow{2}*{Sample size}
&\multirow{2}*{$r$}&\multicolumn{3}{c}{TNNLR} & \multicolumn{3}{c}{TNNRLR} & \multicolumn{3}{c}{NCRLR} & \multicolumn{3}{c}{NCRRLR}\\\cline{3-14}
  & & logErr & Acc &$\hat{r}$&logErr & Acc &$\hat{r}$&logErr & Acc &$\hat{r}$&logErr & Acc &$\hat{r}$\\\hline
\multirow{3}*{2000}&2&0.5536&	81.13&	2.03&	0.5488&	81.43&	2.01&	-0.3950&	81.50&	2.00&	-0.4506&	82.17&	2.00\\
 & 4 &1.0525&	86.04&	4.01&	1.0365	&86.52&	4.03&	0.0592&	87.47&	4.00&	0.0204&	88.76&	4.00\\
  & 8 &1.5170&	83.57&	7.70&	1.4914&	87.27&	8.00&	1.2016&	88.09&	7.90&	0.8657&	92.54&	7.90\\\hline
 \multirow{3}*{4000}&2 &0.4135&	81.14&	2.00&	0.4132&	81.18&	2.00&	-0.4895	&81.35&	2.01&	-0.5918&	81.54&	2.00\\
  &4 &0.9466&	85.49&	4.02&	0.9341&	85.71&	4.01&	-0.0300&	85.97&	4.00&	-0.2250&	86.66&	4.00\\
  &8&1.4400&	88.10&	8.00&	1.4310&	88.85&	8.00	&0.4627	&89.99&	8.00&	0.3160&	91.34&	8.00\\\hline
\multirow{3}*{6000}&2&0.3129&	81.27&	2.00&	0.3102&	81.26&	2.00&	-0.5672	&81.27&	2.00&	-0.7105&	81.42&	2.00\\
 &4&0.8771&	85.71&	4.00&	0.8658&	85.85&	4.00&	-0.0627	&85.85&	4.00&	-0.3077&	86.30&	4.00\\
 &8&1.3904&	88.93&	8.00&	1.3905&	89.30&	8.00&	0.4637&	89.85&	8.00&	0.2904&	90.75&	8.00\\
\hline\hline
\end{tabular}
}
\end{center}
\end{table}

\begin{figure}[!htb]
\centering
\subfigure[]{
		\includegraphics[scale=0.8]{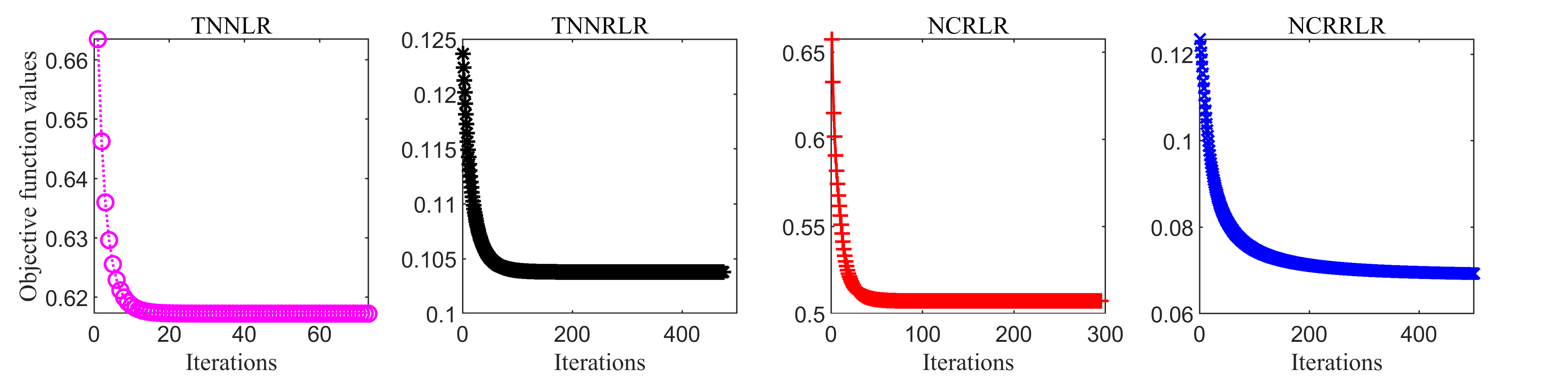}}
\caption{Convergence behavior of logistic regression models.}
\label{Simu-F3}
\end{figure}

\textbf{(b) Effect of robustness:} We first vary the misspecification rate $p_m\%\in\{0,5\%,10\%,15\%\}$ to investigate the robustness of different estimators, and then some corrupted entries is added to predictors. To be specific, there are $p_n\%$ predictors have $p_c\%$ elements contaminated by the Pareto distribution, where $p_n$ and $p_c$ is taken from $0,5,10$ to $15$. Such setting is common in robust matrix/tensor recovery and image classification. We fixedly set $n=2000$, $r=2$, and the dimension is taken as above. It is clear that the estimation error and prediction accuracy degrades as the misspecification rate grows, from Table \ref{Simu-T7}. And our proposed NCRRLR achieves the best Err and Acc, and could yields more accurate rank estimation. Notably, when the misspecification rate equals to $0$, NCRRLR could yield the highest prediction accuracy albeit the estimation error is not least.
\begin{table}[hptb]
\begin{center}
\caption{The averaged logErr, Acc and $\hat{r}$ of different misspecification rate.}\label{Simu-T7}
\footnotesize
\resizebox{\textwidth}{!}{
\begin{tabular}{ccccccccccccc}\hline\hline
\multirow{2}*{$p_m$}
&\multicolumn{3}{c}{TNNLR} & \multicolumn{3}{c}{TNNRLR} & \multicolumn{3}{c}{NCRLR} & \multicolumn{3}{c}{NCRRLR}\\\cline{2-13}
  & logErr & Acc &$\hat{r}$&logErr & Acc &$\hat{r}$&logErr & Acc &$\hat{r}$&logErr & Acc &$\hat{r}$\\\hline
0&0.4690&	81.61&	2.00&	0.5082&	81.67&	2.00&	-0.5171&	81.94&	2.00&	-0.4702&	82.57&	2.00\\
5&	0.5536&	81.13&	2.03&	0.5488&	81.43&	2.01&	-0.3950&	81.50&	2.00&	-0.4506&	82.17&	2.00\\
10&	0.6216&	81.09&	2.06&	0.6028&	81.33&	2.01&	-0.0168&	81.39&	2.01&	-0.1689	&82.04&	2.00\\
15&	0.6805&	80.35&	2.02&	0.6539&	80.82&	2.02&	0.2475&	80.57&	2.01&	0.0534	&81.71&	2.00\\
\hline\hline
\end{tabular}
}
\end{center}
\end{table}
From Table \ref{Simu-T8}, when the sample is clean (there has no contaminated point), the naive MLE of LR obtains lower error. However, the robust LR achieve higher prediction accuracy albeit its estimation error is not least. As the corrupted samples increase, the same argument as Table \ref{Simu-T7} holds. The estimation error grows, prediction accuracy deduces, and estimated rank breaks slowly, as $p_n$ and $p_c$ increase. But, it is easy to see that our proposed robust LR is insensitive to the contaminated samples.

\begin{table}[hptb]
\begin{center}
\caption{The averaged logErr, Acc and $\hat{r}$ of different setting.}\label{Simu-T8}
\footnotesize
\resizebox{\textwidth}{!}{
\begin{tabular}{cccccccccccccc}\hline\hline
\multirow{2}*{$p_n$}&\multirow{2}*{$p_c$}
&\multicolumn{3}{c}{TNNLR} & \multicolumn{3}{c}{TNNRLR} & \multicolumn{3}{c}{NCRLR} & \multicolumn{3}{c}{NCRRLR}\\\cline{3-14}
 & & logErr & Acc &$\hat{r}$&logErr & Acc &$\hat{r}$&logErr & Acc &$\hat{r}$&logErr & Acc &$\hat{r}$\\\hline
0&	0&	0.4661&	81.67&	2.00&	0.4706&	81.85&	2.00&	-0.4743	&81.89&	2.00&	-0.4638&	82.66&	2.00\\\hline
\multirow{3}*{5}&5&	0.7036&	80.54&	2.10&	0.5949&	81.43	&2.00&	0.5713&	80.76	&2.10&	-0.4362&	82.22&	2.00\\
 &10	&0.7232	&80.67	&2.00&	0.6114&	81.41	&2.10&	0.6238&	80.61&	2.00&	-0.3697	&82.29&	2.00\\
 &15&	0.7908&	78.11&	2.08&	0.7144&	81.00&	2.10&	0.6949&	79.78&	2.07&	-0.3365&	82.17&	2.00\\\hline
\multirow{3}*{10}&5	&0.7260	&80.67&	2.00	&0.6184&	81.43&	2.00&	0.5072&	80.78&	2.00&	-0.4022&	81.97&	2.00\\
 &10&	0.7348&	77.73&	2.50&	0.6575&	80.58&	2.70&	0.5740&	79.45&	2.18&	-0.1790&	81.22&	2.10\\
 &15&	0.8627&	68.35&	2.80&	0.8547	&76.55&	2.60&	0.7804&	71.90&	2.40	&-0.0998&	81.35&	2.16\\\hline
\multirow{3}*{15}&5&	0.7496&	79.40	&2.00&	0.6526&	80.50&	2.20&	0.6152&	79.75&	2.00&	-0.2015&	81.52&	2.00\\
 &10&0.8147&	78.75&	2.20&	0.7300&	79.90&	2.15&	0.7459&	79.55&	2.10&	-0.1356	&81.56&	2.00\\
  &15&0.8532&	70.25&	3.10&	0.8047&	75.45&	2.80&	0.8215&	71.75&	3.05&	-0.0738	&81.25	&2.07\\
\hline\hline
\end{tabular}
}
\end{center}
\end{table}

\section{Discussion}\label{Sec6}
In this paper, we have developed a unified framework of tensor regression where we allow that both the loss and penalty functions are nonconvex. Although our model can be entirely nonconvex, with some mild conditions, the proposed estimation algorithm is easy to implement, and enjoys global convergence with at least sub-linear algorithmic convergence rate. Moreover, under some regularity assumptions, the desired statistical properties are also established. It suggests that all stationary points of proposed nonconvex problem are statistically consistent. A series of numerical studies and applications confirm our theoretical results.

The direct extension of this research is considering other nonconvex loss and penalty functions not included by our present results, such as weighted t-TNN and weighted Schatten $p$-norm. In addition, in the case of big data, more and more data are stored in the distributed setting. Therefore, it is necessary to extend the nonconvex models to  distributed scenario without leaking privacy. However, it may be need more powerful theoretical tools to establish statistical properties. Finally, it would be interesting to develop nonconvex model under semi-supervised framework since there exists numerous unlabeled data and labeling data is expensive.


\bibliographystyle{dcu}
\bibliography{paper,book}

\end{document}